%% file: feedbackMACv26.tex
\documentclass[final,onecolumn]{IEEEtran}

\input{preamble}
\usepackage{dsfont}
\usepackage{xspace}
\usepackage{xcolor}

\usepackage[ colorlinks = true,
             linkcolor = blue,
             urlcolor  = blue,
             citecolor = red,
             anchorcolor = green,
]{hyperref}

\usepackage{cite}
\allowdisplaybreaks[1]
\begin{document}


\title{On Gaussian Channels with Feedback under   Expected Power Constraints  and  with Non-Vanishing Error Probabilities }

\author{Lan V.\ Truong  $\quad$ Silas L.\ Fong $\quad$
        Vincent Y.~F.~Tan
\thanks{The authors  (Emails: lantruong@u.nus.edu, silas\_fong@nus.edu.sg, vtan@nus.edu.sg) are with the  Department of Electrical and Computer Engineering, National University of Singapore (NUS). V.~Y.~F.~Tan is also with the Department of Mathematics, NUS. }
 \thanks{The authors are supported by an NUS grant (R-263-000-B37-113) and  by  a Singapore Ministry of Education (MOE) Tier 2 grant (R-263-000-B61-112).} \thanks{This paper was presented in part at SPCOM 2016~\cite{TFT16_spcom} and ISIT 2016~\cite{TFT16}.}
}

\maketitle

\begin{abstract}
In this paper, we consider single- and multi-user Gaussian channels with feedback under expected power constraints and with non-vanishing error probabilities. In the first of two  contributions, we study asymptotic expansions for the additive white Gaussian noise (AWGN) channel with feedback under  the average error probability formalism. By drawing ideas from Gallager and Nakibo\u{g}lu's work  for the direct part and the meta-converse for the converse part, we establish the $\eps$-capacity and show that it  depends on $\eps$ in general and so the strong converse fails to hold. Furthermore, we provide bounds on the second-order term in the asymptotic expansion. We show that for any positive integer $L$,  the second-order term is bounded between a term proportional to $-\ln_{(L)} n$ (where $\ln_{(L)}(\cdot)$ is the $L$-fold nested logarithm function) and a term proportional to $+\sqrt{n\ln n}$ where $n$ is the blocklength. The lower bound on the second-order term shows that feedback does provide an improvement in the maximal achievable rate over the case where no feedback is available. In our second contribution, we establish the $\eps$-capacity region for the AWGN multiple access channel (MAC) with feedback under the expected power constraint by combining ideas from hypothesis testing, information spectrum analysis, Ozarow's coding scheme, and power control.
\end{abstract}
\section{Introduction}
Shannon showed that feedback does not increase the capacity of a discrete memoryless channel (DMC) \cite{Sha56}. It is known, however, that  feedback can improve the error probability performance \cite{Burnashev} and also simplify coding schemes \cite{SK66}. As an example, Polyanskiy, Poor and Verd\'u showed that  variable-length feedback \cite{PPV11b} improves on the capacity term if a non-vanishing error probability is allowed. Altu\u{g} and Wagner \cite{AW14} showed that  full-output feedback   can, in some scenarios, improve on the second-order asymptotics of a DMC even if the first-order capacity term is not improved. Although there has been some progress to evaluate the second-order asymptotics  for the DMC or the AWGN channel with fixed-length full-output feedback, all the results derived thus far are mainly for weakly input-symmetric DMCs \cite{PPV11b} or the AWGN channel under peak power constraint \cite{FT15}, where the distribution of the relevant information density  (between the channel and the capacity-achieving output distribution) is invariant to all channel input symbols that undergo a random transformation. See \cite[Sec.~V.A]{PPV11b} for details. 

In this paper, we provide  two main contributions. First, we derive the  so-called {\em $\eps$-capacity} of the AWGN channel with full-output feedback~\cite{ShaFeder11, SK66, Ozarow}  under an expected power constraint and the average error probability formalism. The $\eps$-capacity is  the supremum of all rates for which it is guaranteed that there exists a sequence of codes whose asymptotic error probability is upper bounded by $\eps$. By expected power constraint, we mean that for a given message set $\calW$, all the inputs to the channel $\{ x_k(w,Y^{k-1}): k=1,\ldots, n , w \in\calW\}$ must satisfy
\begin{equation}
\frac{1}{ |\calW| } \sum_{w\in\calW} \frac{1}{n}\sum_{k=1}^n \bbE[x_k^2(w,Y^{k-1})] \le P\label{eqn:expected_pow}
\end{equation}
for some constant power $P>0$.  Notice that in addition to averaging over the time slots $k= 1,\ldots, n$, we average over the entire codebook (or messages in the message set $\calW$). The latter averaging is also known in the wireless communications community as a {\em long-term power constraint}~\cite{caire99, yang15}.  The capacity of this channel is clearly
\begin{equation}
\rvC(P) = \frac{1}{2}\ln(1+P), \quad \mbox{nats per channel use}. \label{eqn:gauss_cap}
\end{equation}
This is the maximal achievable rate when the error probability is required to vanish. If the average error probability does not vanish and is bounded above by some $\eps\in[0,1)$ asymptotically, the corresponding quantity one seeks is the  {\em $\eps$-capacity}.

In the second contribution, we establish the {\em  $\eps$-capacity region} for the AWGN-MAC with feedback under an expected power constraint similar to \eqref{eqn:expected_pow}. The capacity region in which the error probability is required to vanish was established by Ozarow~\cite{Ozarow}. We generalize Ozarow's seminal capacity result to the case where the error probability need not vanish.  Our investigations are motivated by the recent interest in the practicality of establishing finite blocklength fundamental limits~\cite{PPV10} of various channel and source models. Finding the $\eps$-capacity region is a first step in making progress to understanding the non-asymptotic fundamental limits of any network channel model. Estimating the second-order behavior provides a refinement to this understanding.

\subsection{Elaborations on the Main Contributions} \label{sec:elab}
We now elaborate on our two main contributions.

 First, by combining the posterior matching~\cite{ShaFeder11} arguments by Schalkwijk and  Kailath (SK)~\cite{SK66}, Ozarow~\cite{Ozarow}, Gallager and Nakibo\u{g}lu~\cite{GalN} and Ihara~\cite{Ihara12} with a power control argument~\cite{yang15}, we show that the $\eps$-capacity under the constraint in  \eqref{eqn:expected_pow} is $\rvC(P/(1-\eps))$ and so the strong converse does not hold (as is expected).  Nevertheless, the $\eps$-capacity is unchanged as compared to the case without feedback~\cite{yang15} so feedback apparently does not help to improve (increase) the first-order term. One then wonders about the effect of feedback on the second- and higher-order asymptotics~\cite{TanBook,PPV10,Hayashi09}. 
We show that under the constraint in \eqref{eqn:expected_pow},  for all positive integers $L$, the maximum number of messages transmissible through $n$ uses of the AWGN channel with average error probability no larger than $\eps$, namely $M^*_{\mathrm{fb}}(n,P,\eps)  $, satisfies
\begin{align}
n\rvC\left(\frac{P}{1-\eps}\right)  -O\left(\ln_{(L)}(n)\right) +O(1) &\le \ln M^*_{\mathrm{fb}}(n,P,\eps)  \label{eqn:lower_bd}\\
& \le
n\rvC\left(\frac{P}{1-\eps}\right) + B_{\eps}\sqrt{n\ln n}+ O(\sqrt{n}),\label{eqn:upper_bd}
\end{align}
where $\ln_{(L)}(\cdot)$ is the $L$-fold nested logarithm function (defined in \eqref{eqn:mult_log} to follow) and $B_{\eps} >0$ is some positive constant defined in \eqref{eqn:def_B_eps} in the sequel. See Table \ref{tab:Comparsion}. As we shall see, the implication of the lower bound in \eqref{eqn:lower_bd} is that feedback greatly improves the second-order term in the asymptotic expansion under the expected power constraint, compared to the no feedback case in the analogous long-term power constraint where the codewords $\{x^n(w): w\in\calW\}$ are required to satisfy
\begin{equation}
 \frac{1}{ |\calW| } \sum_{w\in\calW} \frac{1}{n}\sum_{k=1}^n x_k^2(w ) \le P \label{eqn:lt_pow} .
\end{equation}
To obtain the nested logarithm in the lower bound in \eqref{eqn:lower_bd}, we appeal to a modification of the  SK coding scheme~\cite{SK66} by  Gallager and Nakibo\u{g}lu~\cite{GalN} and Ihara~\cite{Ihara12} that guarantees that for all rates below capacity, the probability of error for an AWGN channel with feedback decays multiply-exponentially fast.  Also see the works by Pinsker~\cite{Pinsker68}, Kramer~\cite{Kramer69} and Zigangirov~\cite{Zigangirov} which all show that for fixed rates below capacity, the error probability decreases faster than an exponential
of any order.

Second, we generalize and strengthen a seminal result by Ozarow~\cite{Ozarow} concerning AWGN-MACs with feedback under an expected power constraint. In his seminal paper, Ozarow showed that the capacity region of the AWGN-MAC with feedback is the set of rate pairs $(R_1, R_2)$ such that
\begin{align}
R_1 &\le\rvC( (1-\rho^2 ) P_1) \label{eqn:macfb1} \\
R_2 &\le\rvC( (1-\rho^2 ) P_2) \\
R_1+R_2 &\le\rvC( P_1+P_2 + 2\rho\sqrt{P_1 P_2}  ) \label{eqn:macfb3}
\end{align}
for some $\rho \in [0,1]$ and
where $P_j$ (for $j = 1,2$) is the signal-to-noise ratio of receiver $j$. If one allows the average error probability to be non-vanishing, say bounded above by $\eps\in [0,1)$, then one wonders whether the region  in \eqref{eqn:macfb1}--\eqref{eqn:macfb3} is enlarged and if so, by how much? We provide a complete answer to this question and show that the signal-to-noise ratios $P_j$ are modified to be $P_j / (1-\eps)$. We also provide bounds on the second-order terms. The techniques in this paper  leverage several   ideas from the literature including the meta-converse~\cite{PPV10}, information spectrum analysis \cite{Han10} for channels with feedback~\cite{Alajaji95, ChenAlajaji95}   and Ozarow's achievability and weak converse~\cite{Ozarow}. In particular, an important {\em ``single-letterization'' lemma} (Lemma \ref{mac-lem6}) is developed to introduce the single parameter $\rho$ in \eqref{eqn:macfb1}--\eqref{eqn:macfb3}  in order to facilitate the outer bounding. This non-standard lemma forms the crux of our converse proof.

\subsection{Related Works}
\begin{table}
	\centering
\begin{tabular}{|c|c|c|c|c|} 
\hline
\multicolumn{5}{|c|}{\textbf{Second-Order Term}} \\
\hline
\multirow{2}{*}{}&\multicolumn{2}{|c|}{Average Error Probability}&\multicolumn{2}{|c|}{Maximum Error Probability} \\ \cline{2-5}
 & No  Feedback  & Feedback  & No  Feedback & Feedback \\ 
\hline
$   \! $  Peak Power $ \!   $   & $A_{\eps}\sqrt{n}$  & $A_{\eps}\sqrt{n}$ & $A_{\eps}\sqrt{n}$& $A_{\eps}\sqrt{n}$\\\hline
References &\cite{Hayashi09, PPV10} &$    $ \cite{Sha59b,Pinsker68,FT15} $    $& \cite{Hayashi09, PPV10}  &$    $\cite{Sha59b,Pinsker68,FT15} $  \!  $ \\
\hline
$  \!  $  Expected Power $  \!  $   & $    -B_{\eps}\sqrt{n\ln n}    $ & $  \!   $ $[-O(\ln_{(L) } n) , B_{\eps}\sqrt{n\ln n}]$ $  \!   $  & $A_{\eps}\sqrt{n}$ & $   \! $ $[-O(\ln_{(L)} n) , B_{\eps}\sqrt{n\ln n}]$  $ \!   $  \\\hline 
References & \cite{yang15}& Theorem \ref{thm:main} & $  $\cite[Thm.~73]{Pol10} $  $ & Theorem~\ref{thm:main} \\\hline
\end{tabular}
\vspace{2mm}
	\caption{Second-order terms in asymptotic expansions for the AWGN channel under different constraints.  Note that when the second-order term is $A_\eps\sqrt{n}$ (resp.\ $-B_\eps\sqrt{n\ln n}$), the first-order term is $n\rvC(P)$ (resp.\  $n\rvC(P / (1-\eps))$). }
	\label{tab:Comparsion}
\end{table}

\begin{figure}
\centering
\includegraphics[width =.4\columnwidth]{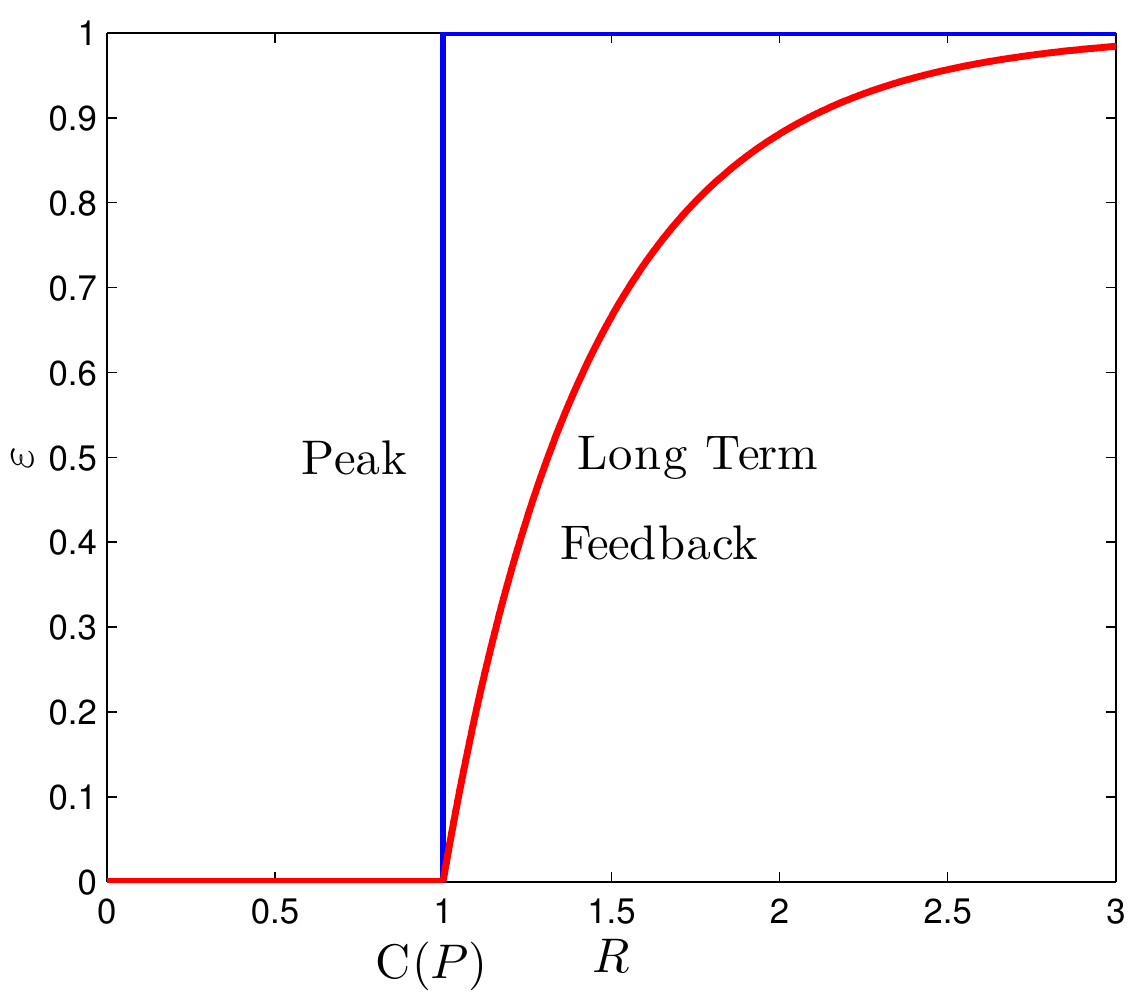}
\caption{The behavior of the optimal asymptotic error probability as a function of the rate under the peak \cite{FT15} and the long-term power constraints in~\eqref{eqn:lt_pow}  as well as with feedback in~\eqref{eqn:expected_pow}. See  Theorem \ref{thm:main}. }\label{fig:peak_lt}
\end{figure}

Table \ref{tab:Comparsion}   lists all the previous results and references for the second-order term in the asymptotic expansions for AWGN channels with and without feedback under both error formalisms. In the table,  we use the abbreviations
\begin{align}
A_{\eps}:=\sqrt{\rvV(P)}\Phi^{-1}(\eps)\quad\mbox{and}\quad
B_{\eps}:=\sqrt{\rvV\left(\frac{P}{1-\eps}\right)}, \label{eqn:def_B_eps}
\end{align}
where the {\em Gaussian dispersion function} is
\begin{align}
\rvV(x):=\frac{x(x+2)}{2(x+1)^2} \hspace{2mm} \mbox{nats$^2$ per channel use} \label{eqn:gauss_disp}
\end{align}
and the {\em Gaussian cumulative distribution function (cdf)} is
\begin{align}
\Phi(u):=\frac{1}{\sqrt{2\pi}}\int_{-\infty}^u \exp\left(-\frac{t^2}{2}\right) \,\rmd t.
\end{align}
We note that when the second-order term is $A_\eps\sqrt{n}$, the first-order term is $n\rvC(P)$. In the case where the second-order term is $-B_\eps\sqrt{n\ln n}$, the first-order term is $n\rvC(P / (1-\eps))$. Also see Fig.~\ref{fig:peak_lt}.
In some scenarios, the third-order term may be  determined. For example, under the average error probability formalism in the no feedback case and peak power constraints, the third-order term is known to be $\frac{1}{2}\ln n$ \cite{TanTom13a} and \cite[Thm.~73]{Pol10}.

Note that for the AWGN channel without feedback under  a long-term (expected) power constraint in \eqref{eqn:lt_pow}, Yang {\em et al.}~\cite{yang15} showed that
\begin{align}
 \ln M^*(n,P,\eps)= n \rvC\left(\frac{P}{1-\eps}\right)-B_\eps\sqrt{n\ln n}\!+\! o\left(\sqrt{n}\right).
\end{align}
Comparing this to \eqref{eqn:lower_bd}, we see that even though the first-order term is unchanged, the achievable second-order term have derived is much improved in the presence of full-output noiseless feedback.  In particular, the backoff proportional to $\sqrt{n \ln n}$ is replaced by a backoff of $\ln_{(L)} (n)$ as $n$ grows.

For the Gaussian MAC with feedback, to the best of the authors' knowledge, there has been no work that attempts to establish the $\eps$-capacity region or the second-order asymptotics. Without feedback, inner bounds for the second-order coding rates for the Gaussian MAC were independently  established by Scarlett, Martinez, and Guill\'en i F\`abregas~\cite{Scarlett15} and MolavianJazi and Laneman~\cite{Mol13}.  The strong converse, together with (non-matching) outer bounds for the second-order coding rates, was established by Fong and Tan~\cite{FongTan15}. For the Gaussian MAC with degraded message sets, the complete second-order asymptotics was derived by Scarlett and Tan~\cite{ST15}.
\subsection{Paper Organization}
The rest of this paper is organized as follows: We state and prove our results concerning the AWGN channel with feedback in Section \ref{sec:awgn}. We do the same for the AWGN-MAC with feedback in Section~\ref{sec:mac}. We conclude our discussion in Section \ref{sec:concl}. Proofs that are more technical are deferred to the appendices.
\section{AWGN Channel with Feedback} \label{sec:awgn}
\subsection{Notation, Channel Model,  and Definitions}

\subsubsection{Notation} We use $\ln x$ to denote the natural logarithm so information units throughout are in nats.  We set $\ln^+(x) :=\ln x$ for $x>0$ and $\ln^+(x):=-\infty$ if $x\leq 0$.  Define    
\begin{equation}
\exp_{(k)}(t):=\underbrace{\exp(\exp(\ldots\exp(t)))}_{\text{$k$ times}},\quad\forall\, t\in\bbR
 \end{equation}
 to be the {\em multiple (nested)  exponential function}.
For any $L\in\mathbb{N}$ and every $k=1, 2, \ldots, L$, we define the {\em multiple (nested)  logarithm function} $\ln_{(k)}(n)$ for every $n\ge \exp_{(L)}(1)$ in a recursive way as follows:
\begin{align}
\ln_{(k)}(n):= \begin{cases}
\ln n & \text{if $k=1$,}\\
 \ln\ln_{(k-1)} (n) & \text{otherwise.}
 \end{cases} \label{eqn:mult_log}
\end{align}
 Random variables and information-theoretic quantities are standard and mainly follow the text by El Gamal and Kim~\cite{elgamal}. We also use asymptotic notation such as $O(\cdot)$ in the standard manner; $f(n)=O(g(n))$ holds if and only if the implied constant $\limsup_{n\to\infty} |f(n)/g(n)|<\infty$.

\subsubsection{Channel Model}
 We consider the standard AWGN  channel model
\begin{equation}
 Y_k=X_k+Z_k, \qquad k = 1,\ldots, n
 \end{equation}
 where $Z_k$ is independent and identically distributed Gaussian noise with zero mean and unit variance.   Thus, for a single-channel use,  the channel from $X$ to $Y$ can be written as
\begin{align}
V(y|x )=\frac{1}{\sqrt{2\pi}}\exp\left(-\frac{1}{2}(y-x )^2\right).
\end{align}
The channel is used $n$ times in a memoryless channel with feedback.  The input to the channel $X^n=(X_1,\ldots, X_n)$ is power constrained (to be stated precisely in \eqref{eqn:exp_pow}  in Definition \ref{def:awgn}).

\subsubsection{Basic Definitions} Now we state some important definitions for the AWGN channel with feedback. Please refer to \cite[Fig.~3.4]{elgamal} for an illustration of the setup of the problem.
\begin{definition}\label{def:awgn}
An $(n,M,P)$-feedback code under  the expected power constraint consists of the following:
\begin{itemize}
\item A message set
\begin{align}
\mathcal{W}:=\{1,2,...,M\}.
\end{align}
Message $W$ is uniformly distributed on $\mathcal{W}$.
\item A collection of encoding functions
\begin{align}
f_k: \mathcal{W}\times \mathbb{R}^{k-1}\to \bbR
\end{align}
for each $k \in \{1,2,...,n\}$, where $f_k$ is the encoding function at node $s$ for encoding $X_k$ such that
\begin{align}
X_k=f_k(W,Y^{k-1})
\end{align}
and
\begin{align}
\frac{1}{n}\sum_{k=1}^n \bbE[X_k^2] \leq P.\label{eqn:exp_pow}
\end{align}
\item A decoding function
\begin{align}
\phi_n: \mathbb{R}^n \to \mathcal{W},
\end{align}
where $\mathcal{W}$ is the decoding function for $W$  such that
\begin{align}
\hat{W}=\phi_n(Y^n).
\end{align}
\end{itemize}
\end{definition}
The expectation in \eqref{eqn:exp_pow} is over both the message $W$ and the channel outputs $Y^{k-1}$, thus \eqref{eqn:exp_pow} is identical to   \eqref{eqn:expected_pow}.

\begin{definition}
For an $(n,M,P)$-feedback code defined on AWGN channel with feedback, we can calculate the average probability of error
\begin{equation}
P_{\mathrm{avg}}:=\Pr(\hat{W}\neq W).
\end{equation}
  We call an $(n,M,P)$-feedback code with average probability of decoding error no larger than $\eps$   an {\em $(n,M,P,\eps)$-feedback code}. We may also define the maximum number of messages
\begin{align}
M_{\mathrm{\mathrm{fb}}}^*(n,P,\eps):=\max\left \{M: \exists\hspace{1mm}\mbox{an $(n,M,P,\eps)$-feedback code}\right \}. \label{eqn:Mstar}
\end{align}
Similarly, we can calculate the maximal probability of error
\begin{equation}
P_{\mathrm{max}}:=\max_{w\in\mathcal{W}}\Pr(\hat{W}\neq w|W=w). \label{maxError}
\end{equation}
\end{definition}
The average probability of error in the above definition is defined over the randomness of the message $W$,  the channel outputs $Y^{k-1}, k = 1,\ldots, n$ (in the expected power constraint in \eqref{eqn:expected_pow}) and the channel noise $Z^n\sim \calN(0_n,I_{n\times n})$. Naturally, for the maximum probability of error in \eqref{maxError}, we may also define $M_{\mathrm{\mathrm{fb}}}^*$ similarly to \eqref{eqn:Mstar}. We note, however (cf.\ Table \ref{tab:Comparsion}), that it will become apparent that  the results do not depend on whether the maximum or average error probability formalism is employed so we do not distinguish between $M_{\mathrm{\mathrm{fb}}}^*$ for both error probability formalisms.

\subsection{Main Result} \label{subsectionMainResult}
Our main contribution in this section  is the following theorem.
\begin{theorem} \label{thm:main}
Let $\rvC(\cdot )$ be the Gaussian capacity function defined in \eqref{eqn:gauss_cap} and recall the definition of  $B_\eps$  in \eqref{eqn:def_B_eps}. For  any $0<\eps<1$ and  any $L\in\bbN$, the following expressions for the AWGN channel with feedback subject to an expected power constraint hold
\begin{align}
\ln M^*_{ \mathrm{fb}}(n,P,\eps)& \geq n\rvC\left(\frac{P}{1-\eps}\right)-2^{L-2}\ln_{(L)}( n) +O(1), \label{eqn:dire} \\
 \ln M^*_{\mathrm{fb}}(n,P,\eps) &\leq n\rvC\left(\frac{P}{1-\eps}\right) + B_{\eps} \sqrt{n\ln n}+O(\sqrt{n}).\label{eqn:converse_su}
\end{align}
These bounds imply  that the $\eps$-capacity
\begin{equation}
\lim_{n\to\infty}\frac{1}{n}  \ln M^*_{\mathrm{fb}}(n,P,\eps)=\rvC\left(\frac{P}{1-\eps}\right). \label{eqn:no_str_conv}
\end{equation}

\end{theorem}
The achievability and converse parts are proved in Sections~\ref{section1} and \ref{sec:converse} respectively.  The following remarks are now in order.

First, \eqref{eqn:no_str_conv} means that the strong converse does not hold for the AWGN channel with feedback under the expected power constraint. Second, the implication of Theorem~\ref{thm:main} is that the first-order term $\rvC\left({P}/{(1-\eps)}\right)$ does not improve when feedback is present  but the second-order term, which is at least $-2^{L-2}\ln_{(L)}( n) n+O(1)$ as shown in \eqref{eqn:dire}, does improve   (cf.\ \cite{yang15} and Table~\ref{tab:Comparsion}).  We note that one is free to choose the value of $L$ in  \eqref{eqn:dire} but the implied constant in the $O(1)$ term depends on~$L$. Third, our achievability also holds under the maximal error formalism (cf.\ \eqref{maxError} and Table~\ref{tab:Comparsion}), because the encoder can relabel the message according to the uniform distribution.
\subsection{Proof of Achievability}\label{section1}
\begin{proposition} \label{prop:ach}
For an AWGN channel with feedback subject to an expected power constraint $P$, the maximum number of transmissible messages $M^*_{\mathrm{fb}}(n,P,\eps)$ satisfies \eqref{eqn:dire}.
\end{proposition}
\begin{IEEEproof}
To prove this proposition, we show that for each $n$, there exists a coding scheme such that the aforementioned expression holds true. Specifically,  we show that a combination of  the two-phase coding scheme \cite{Ihara12} and power control ideas \cite[Sec.\ 4.3.3]{Pol10} can provide the required achievable asymptotic expansion.
\subsubsection{Coding Scheme}
For each fixed finite $n$, we choose\footnote{We ignore integer constraints on the number of codewords $M_n$ in \eqref{eqn:defM_n}. We simply
set $M_n$ to the nearest integer to the number on the right-hand-side of~\eqref{eqn:defM_n}.}
\begin{align}
M_n= \left(\frac{1-\eps_n}{1-\eps}\right) \left(1+\frac{P}{1-\eps+\frac{1}{n}}\right)^{n/2}, \label{eqn:defM_n}
\end{align}
for some $\eps_n< \eps$ to be determined later. In fact, we will choose the parameters of our code so that $\eps_n  \le\frac{1}{n}$ so this constraint ($\eps_n< \eps$) is clearly satisfied for $n$ large enough. See \eqref{eqn:Pavg_GN}  to follow. Note that the ratio $\frac{1-\eps_n}{1-\eps}=\Theta(1)$, which is essential in the arguments to follow.    We  perform the following tasks:
\begin{itemize}
\item We divide the set of $M_n$ messages into two subsets. The first subset consists of
\begin{align}
\overline{M}_{n}=\left(\frac{1-\eps}{1-\eps_n}\right) M_n=\left(1+\frac{P}{1-\eps+\frac{1}{n}}\right)^{n/2}  \label{eqn:def_M1n}
\end{align}
messages called $\calA_1\subset\calW$ and the second subset consists of $ M_n - \overline{M}_{n}$ messages  called $\calA_2=\calW\setminus\calA_1$.
\item For messages $w$ in the first message subset $\calA_1$ we use the Gallager-Nakibo\u{g}lu (GN) two-phase coding scheme~\cite{GalN} (which is itself based on  the Schalkwijk-Kailath~\cite{SK66} coding scheme and a result of Elias \cite{Elias56} concerning the minimum mean-square
distortion achievable in transmitting a single Gaussian random
variable over multiple uses of the same Gaussian channel)  and transmit all ``codewords'' $\{X_k(w,Y^{k-1} ) : k = 1,\ldots, n\}$ with expected power less than or equal to ${P}/{(1-\eps+1/n) }$ with average error probability that is bounded above by some $\eps_n$ (to be computed).
\item For messages $w$ in the second message subset $\calA_2$, we encode all of them by the all-zero codeword.
\end{itemize}
\subsubsection{Analysis}
By the above feedback coding scheme, it is obvious that the expected transmission power is
\begin{align}
P_{\mathrm{avg}}\leq \frac{\overline{M}_{n}}{M_n}\left(\frac{P}{1-\eps+\frac{1}{n}}\right) = \left(\frac{1-\eps}{1-\eps_n}\right) \left(\frac{P}{1-\eps+\frac{1}{n}}\right).
\label{eqn:Pavg}
\end{align}
Since the messages are uniformly distributed, the overall average error probability of the proposed coding scheme (combination of Gallager-Nakibo\u{g}lu~\cite{GalN, Ihara12} and a power control idea~\cite{Pol10}) is upper bounded as
\begin{align}
P_\rme^{(n)} \leq \eps_n\cdot\frac{\overline{M}_{n}}{M_n}+1\cdot\frac{M_n-\overline{M}_{n}}{M_n} =\eps.
\end{align}
Now, we will show that the error probability of the GN scheme $\eps_n=P_{\mathrm{GN}}^{(n)}$ is upper bounded by $1/n$, the average power $ P_{\mathrm{avg}}$ is upper bounded by $P$, and the maximum number of messages $M_n$ that can be transmitted through the channel satisfies $\ln M_n =n\rvC\left( \frac{P}{1-\eps}\right)-O\big(\ln_{(L)} (n)\big) $ for any fixed $L\in\mathbb{N}$.

\paragraph{GN-Scheme Error Probability Analysis}
Fix an $L\in\mathbb{N}$ and $n\ge \exp_{(L)}(1)$ (so $\ln_{(L)}(n)$ is well-defined per \eqref{eqn:mult_log}). Define
\begin{equation}
\delta_n := \frac{L}{n},
\end{equation}
choose $n_1:=\lfloor(1-\delta_n)n\rfloor +1$. Let the rate of the code   be
 \begin{align}
R_{n}&:=\frac{1}{n} {\ln \overline{M}_{n}}
\label{eq31:newest}\\
&=\rvC\left(\frac{P}{1-\eps+1/n}\right)(1-\delta_n)- \frac{2^{L}\ln_{ (L+1) }(n)}{2n }. \label{eqn:Rn2}
\end{align}
In addition, noting that
\begin{equation}
n-n_1+1 = n-\lfloor(1-\delta_n)n\rfloor \in \{L, L+1\},
\end{equation}
 we define for each $k=0,1,2,\ldots,n-n_1+1$,
\begin{align}
D_k&:=
\frac{\ln_{(L+1)}(n^{3^{L-k-1}})}{2n}
\end{align}
and
\begin{align}
\tilde D_k&:=
\frac{\ln_{(L+1)}(n^{3^{L-k}/2})}{2n}
\end{align}
such that
\begin{equation}
0<\tilde D_{n-n_1+1}<D_{n-n_1}<\tilde D_{n-n_1}<\cdots < D_0<\tilde D_0<\rvC\left(\frac{P}{1-\eps+1/n}\right)(1-\delta_n)-R_n = \frac{2^{L}\ln_{ (L+1) }(n)}{2n }.
\end{equation}At this point, we leverage 
a useful non-asymptotic estimate on the average error probability of the GN scheme provided by Ihara \cite[Theorem~1]{Ihara12} who analyzed the error probability for Gaussian channels with stationary but possibly non-memoryless (non-white) feedback. This estimate says that  for~$n$ sufficiently large ($D_k$ and $\tilde D_k$ have been carefully chosen so that \cite[Eq.~(25)]{Ihara12} and the last two chains of inequalities in \cite[proof of Theorem~1]{Ihara12} are satisfied), the average probability of error of the  GN scheme $P_{\mathrm{GN}}^{(n)} $ satisfies
\begin{align}
P_{\mathrm{GN}}^{(n)}& \leq \exp\big(- \exp_{(n-n_1+1)}(2D_{n-n_1-1}n)\big) \\
& \leq \exp\big(- \exp_{(L)}(2D_{L-1}n)\big)\\
& = e^{-\ln n}\\
&=\frac{1}{n}. \label{eqn:Pavg_GN}
\end{align}

\paragraph{Power Consumption Analysis}
It follows from \eqref{eqn:Pavg_GN} that
\begin{align}
P_{\mathrm{avg}}=\frac{P(1-\eps)}{1-\eps+\frac{1}{n} -\eps_n \left(1-\eps+\frac{1}{n}\right)}\leq P,
\end{align}
since we have
\begin{align}
 \eps_n \left(1 -\eps + \frac{1}{n}\right) =\frac{1}{n} \left(1 -\eps + \frac{1}{n}\right)  \leq \frac{1}{n}.
\end{align}

\paragraph{Message Set Size Analysis}
From  \eqref{eqn:def_M1n}, \eqref{eqn:Pavg} and \eqref{eqn:Rn2},  we have for all sufficiently large~$n$
\begin{align}
\ln M_n&= \ln \overline{M}_{n} +\ln\left(\frac{1-\eps_n}{1-\eps}\right) \\
&=n\rvC\left( \frac{P}{1-\eps+1/n}\right) (1-\delta_n)-  2^{L-1}\ln_{ (L+1) }(n)+ \ln\left(\frac{1-\eps_n}{1-\eps}\right) \\
&=n\rvC\left( \frac{P}{1-\eps+1/n}\right) (1-\delta_n)- 2^{L-1}\ln_{ (L+1) }(n)+ O(1).
\end{align}
%
Recalling that $\delta_n= \frac{L}{n}$, we have
\begin{align}
\ln M_n &= n\rvC\left( \frac{P}{1-\eps}\right)-2^{L-1}\ln_{ (L+1) }(n) +O(1).\label{eqn:msg_set_size1}
\end{align}
This completes the proof of the direct part by relabeling $L+1$ as $L$.
\end{IEEEproof}
\begin{remark}
In the conference version of our paper~\cite{TFT16}, the achievable second-order term in \eqref{eqn:dire}  was stated to be $-\ln\ln n+O(1)$.  We used the standard  SK  coding scheme \cite{SK66} therein. Here, the achievable second-order term is improved to $-O\big(\ln_{(L)}(n)\big)$ for any $L\in\bbN$. To recover $-\ln\ln n+O(1)$ in~\cite{TFT16}, we simply set $L=1$ in \eqref{eqn:msg_set_size1}. We thank a reviewer for pointing out that the results of \cite{GalN, Ihara12} may be used to improve on the lower bound in \eqref{eqn:dire}. 
\end{remark}

\subsection{Proof of Converse } \label{sec:converse}

\begin{proposition}
For an AWGN channel with feedback and an expected power constraint, and any $\eps \in (0,1)$, we have \eqref{eqn:converse_su}.
\end{proposition}
\begin{IEEEproof} Our converse proof is based on the following observation. If there exists a   code with $M_{\mathrm{fb}}^*(n,P,\eps)$ messages, then we can find another   code  with  the same  number of messages with average error probability upper bounded by $1-\gamma/\sqrt{n}$ for some $\gamma>0$ and satisfying  the following property:
\begin{align}
\sum_{k=1}^n (X_k')^2 \leq \frac{nP}{1-\eps-\frac{\gamma}{\sqrt{n}}}, \qquad\mbox{almost surely}  .  \label{eqn:new_constraint}
\end{align}
Indeed, given  a feedback code  with $M_{\mathrm{fb}}^*(n,P,\eps)$ messages and  with encoders  $X_k=f_k(W,Y^{k-1}), k=1,2,\ldots, n$ and a decoder $\hat{W}=\phi(Y^n)$ under the  expected power constraint
\begin{align}
\sum_{k=1}^n \bbE[X_k^2] \leq nP, \label{eqn:Xk_constr}
\end{align}
we may construct a new feedback   code with the same message size $M_{\mathrm{fb}}^*(n,P,\eps)$ as follows.
\begin{itemize}
\item New encoding functions for each $k=1,2,\ldots, n$, i.e.,
\begin{align}
X_k'
:=f_k(W,Y^{k-1})\bone\left\{\sum_{i=1}^{k-1} f^2_i(W,Y^{i-1}) \leq  \frac{nP}{1-\eps-\frac{\gamma }{\sqrt{n}}}\right\},
\end{align} where $\bone \{\cdot\}$ is the indicator function.
\item The decoding function is kept unchanged, i.e.,
\begin{align}
\hat{W}=\phi(Y^n).
\end{align}
\end{itemize}
Observe that with this new feedback encoding scheme, the  average error probability is upper bounded as
\begin{align}
&\Pr(\hat{W}\neq \phi(Y^n))\nonumber \\*
&\leq\Pr\left(\left\{\hat{W}\ne \phi(Y^n)\right\}\!\cap\!\left\{ \sum_{k=1}^n f^2_k(W,Y^{k-1}) \leq \frac{nP}{1 - \eps-\frac{\gamma} {\sqrt{n}}}\right\}\right)  +\Pr\left(\sum_{k=1}^n f^2_k(W,Y^{k-1}) > \frac{nP}{1-\eps-\frac{\gamma}{\sqrt{n}}}\right)  \\
&=\Pr\left(\left\{\hat{W}\neq \phi(Y^n)\right\}\cap\left\{X_k'=X_k, k=1 ,\ldots ,n\right\}\right)  +\Pr\left(\sum_{k=1}^n f^2_k(W,Y^{k-1}) > \frac{nP}{1-\eps-\gamma/\sqrt{n}}\right)  \\
&\stackrel{(a)}{\leq} \eps + \frac{\sum_{k=1}^n \bbE[X_k^2]}{\frac{nP}{1-\eps-\gamma/\sqrt{n}}  }\\
&\stackrel{(b)}{\leq} \eps + \frac{nP}{\frac{nP}{1-\eps-\gamma/\sqrt{n}}}\\
&=1-\frac{\gamma}{\sqrt{n}}.
\end{align}
Here, $(a)$ follows by Markov's inequality \cite[Prop.~1.1]{TanBook} and $(b)$ follows from the   power constraint of the original feedback code in~\eqref{eqn:Xk_constr}.

Furthermore, it is easy to see that the new feedback code satisfies  the   input (peak) power constraint in \eqref{eqn:new_constraint}.
From the above observation, we can convert the problem of finding an upper bound for the maximum number of messages $M_{\mathrm{fb}}^*(n,P,\eps)$ under the expected power constraint to the problem of finding an upper bound for $M^*_{\mathrm{fb,pp}}\left(n,{P}/{(1-\eps-\gamma/\sqrt{n})},1-{\gamma}/{\sqrt{n}}\right)$ under the {\em peak power constraint} \cite{FT15}. Here $M^*_{\mathrm{fb,pp}}(n,Q,\eta)$ is the maximum number of messages that can be transmitted over $n$ channel uses with peak power $Q>0$ and average (and hence, also maximum) error probability upper bounded by $\eta \in (0,1)$.   Therefore, using the bound in~\cite[Eq.\ (29)]{FT15}, under the expected power constraint (of our setting), we obtain for any $\zeta_n>0$ that
\begin{align}
&\ln M_{\mathrm{fb}}^*(n-1,P,\eps)\nn\\*
&\leq \ln \zeta_n  -\ln \bigg[\Pr\bigg(\sum_{k=1}^n\frac{1}{2(1+q_{n,\eps})}\big(-q_{n,\eps}Z_k^2+2\sqrt{q_{n,\eps}}Z_k+q_{n,\eps}\big)
 <\ln \zeta_n -\frac{n}{2}\ln \left(1+q_{n,\eps}\right)\bigg)-\left(1-\frac{\gamma}{\sqrt{n}}\right) \bigg], \label{eqn:upper_bd_mfb}
\end{align}
where, for brevity, we denote
\begin{align}
q_{n,\eps}:=&\frac{P}{1-\eps-\frac{\gamma}{\sqrt{n}}}.
\end{align}
The bound in \eqref{eqn:upper_bd_mfb} is   an information spectrum-style relaxation~\cite[Lemma 4]{Hayashi03} of the meta-converse~\cite[Section III.E]{PPV10}. Also see Appendix \ref{app:info_spec}.
By the Berry-Essen theorem~\cite{feller} (see also \cite[Thm.~1.6]{TanBook}) for independent and identically distributed random variables, we have for all $a \in \mathbb{R}, n\in \mathbb{N}$ that
\begin{align}
\left|\Pr\left(\frac{1}{\sigma \sqrt{n}}\sum_{k=1}^n \frac{1}{2(1+q_{n,\eps})} \big(-q_{n,\eps} Z_k^2 + 2\sqrt{q_{n,\eps}}Z_k + q_{n,\eps} \big) \leq  a\right) - \Phi(a)  \right|\le  \frac{T}{\sigma^3 \sqrt{n}} \label{eqn:berryEsseen}
\end{align}
where the relevant statistics are
\begin{align}
 \mu& := \bbE\left[\frac{1}{2(1+q_{n,\eps})}\left(-q_{n,\eps} Z_1^2 + 2\sqrt{q_{n,\eps}} Z_1 +q_{n,\eps}\right)\right]=0,\\
 \sigma&:=\sqrt{\var\left[\frac{1}{2(1+q_{n,\eps})}\left(-q_{n,\eps} Z_1^2 + 2\sqrt{q_{n,\eps}} Z_1 +q_{n,\eps}\right)\right]}\nonumber\\*
&=\sqrt{\frac{q_{n,\eps}(q_{n,\eps}+2)}{2(1+q_{n,\eps})^2} }=\sqrt{ \rvV( q_{n,\eps}  ) } , \label{eqn:sigma_def}\\
 T&:=\bbE\left[\left|\frac{1}{2(1+q_{n,\eps})}\left(-q_{n,\eps} Z_1^2 + 2\sqrt{q_{n,\eps}} Z_1 +q_{n,\eps}\right)\right|^3\right] \nn\\*
&\stackrel{(a)}{\leq} \bigg(\frac{1}{2(1+q_{n,\eps})}\Big(q_{n,\eps}(\bbE[Z_1^6])^{1/3} + 2\sqrt{q_{n,\eps}}(\bbE|Z_1|^3)^{1/3}+q_{n,\eps}\Big)\bigg)^3 <\infty.
\end{align}
Here, $(a)$ follows from Minkowski inequality (or triangle inequality) for the $\ell_3$ norm of random variables. This implies by choosing
\begin{align}
a=\Phi^{-1}\left(1-\frac{\gamma}{\sqrt{n}}+\frac{2T}{\sigma^3 \sqrt{n}}\right)
\end{align} that
\begin{align}
&\Pr\left(\frac{1}{\sigma \sqrt{n}}\sum_{k=1}^n \frac{1}{2(1+q_{n,\eps})}\left(-q_{n,\eps} Z_k^2 + 2\sqrt{q_{n,\eps}}Z_k + q_{n,\eps}\right) \leq \Phi^{-1}\left(1-\frac{\gamma}{\sqrt{n}}+\frac{2T}{\sigma^3 \sqrt{n}}\right)\right)>1-\frac{\gamma}{\sqrt{n}}+\frac{T}{\sigma^3 \sqrt{n}}. \label{eqn:inf_spec}
\end{align}
Now, put
\begin{align}
\gamma:=\frac{2T}{\sigma^3}+1.
\end{align}
From \eqref{eqn:inf_spec}, we obtain
\begin{align}
\Pr \left(\frac{1}{\sigma \sqrt{n}}\sum_{k=1}^n \frac{1}{2(1+q_{n,\eps})}\left(-q_{n,\eps} Z_k^2 + 2\sqrt{q_{n,\eps}}Z_k + q_{n,\eps}\right) \leq \Phi^{-1}\Big(1-\frac{\gamma}{\sqrt{n}}+\frac{2T}{\sigma^3 \sqrt{n}}\Big)\right)>1-\frac{1}{\sqrt{n}}-\frac{T}{\sigma^3 \sqrt{n}}.
\end{align}
Finally, we set
\begin{align}
\ln \zeta_n&=n\rvC\left(\frac{P}{1\!-\!\eps\!-\!\frac{\gamma}{\sqrt{n}}} \right)\!+\! \sigma \sqrt{n} \Phi^{-1}\left(1\!-\!\frac{\gamma}{\sqrt{n}}\!+\!\frac{2T}{\sigma^3 \sqrt{n}}\right)\\
&=n\rvC\left(\frac{P}{1-\eps-\frac{\gamma}{\sqrt{n}}} \right)+ \sigma \sqrt{n} \Phi^{-1}\left(1-\frac{1}{\sqrt{n}}\right),
\end{align} then from \eqref{eqn:upper_bd_mfb} we obtain
\begin{align}
&\ln M_{\mathrm{fb}}^*(n-1,P,\eps) \nn\\*
& \le n\rvC \left(\frac{P}{1 -\eps -\frac{\gamma}{\sqrt{n}}} \right)+\sigma \sqrt{n} \Phi^{-1}\left(1-\frac{1}{\sqrt{n}}\right) -\ln\left(1-\frac{1}{\sqrt{n}}-\frac{T}{\sigma^3 \sqrt{n}}-\left(1-\frac{\gamma}{\sqrt{n}}\right)\right)\\
&=\!n\rvC \left(\frac{P}{1-\eps-\frac{\gamma}{\sqrt{n}}} \right) \!+ \!\sigma \sqrt{n} \Phi^{-1}\left(1\!-\!\frac{1}{\sqrt{n}}\right)\!-\!\ln\left(\frac{T}{\sigma^3 \sqrt{n}}\right). \label{eqn:bd_m_star2}
\end{align}
In addition, let
$\tau_n :=\Phi^{-1} \big(1-\frac{1}{\sqrt{n}}\big)$,
then we have
\begin{align}
\frac{1}{\sqrt{n}}=1-\Phi(\tau_n) \stackrel{(a)}{\leq} e^{-\tau_n^2/2}. \label{eqn:cher}
\end{align}
Here, (a) follows from  the Chernoff bound. It follows that
\begin{align}
\tau_n \leq \sqrt{\ln n} . \label{eqn:bd_t}
\end{align}
From \eqref{eqn:bd_m_star2} and \eqref{eqn:bd_t} we obtain
\begin{align}
\ln M_{\mathrm{fb}}^*(n-1,P,\eps)\leq n\rvC \left( \frac{nP}{1-\eps-\frac{\gamma}{\sqrt{n}}}\right) + \sigma \sqrt{n\ln n} +\frac{1}{2}\ln n + O(\sqrt{n}). \label{eqn:bd_m_star}
\end{align}
Using Taylor's expansion,   the definition  of the Gaussian dispersion in~\eqref{eqn:gauss_disp}, of $\sigma$ in~\eqref{eqn:sigma_def}, and the bound in~\eqref{eqn:bd_m_star}, we obtain the converse bound in~\eqref{eqn:converse_su} as desired.
\end{IEEEproof}
\section{AWGN Multiple Access Channel with Feedback}\label{sec:mac}
\subsection{Channel Model and Definitions}\label{subsection31}
\subsubsection{Channel Model}
The channel model is given by
\begin{align}
Y=X_1+X_2 +Z,
\end{align} where $X_1$ and $X_2$ represent the inputs to the channel, $Z \sim \mathcal{N}(0,1)$ is an additive Gaussian noise with  zero mean and unit variance, and $Y$ is the output of the channel. Thus, the channel from $(X_1,X_2)$ to $Y$ can be written as
\begin{align}
W(y|x_1,x_2)=\frac{1}{\sqrt{2\pi}}\exp\left(-\frac{1}{2}(y-x_1-x_2)^2\right). \label{eqn:Wchannel}
\end{align}
The channel is used $n$ times in a memoryless channel with feedback. This means that across a block of length $n$, we have
\begin{align}
Y_k=X_{1k}+ X_{2k}+ Z_k, \qquad  k=1,2\ldots,n,
\end{align} where the $Z_k$'s are independent and standard normal, i.e., $Z_k \sim \mathcal{N}(0,1)$. 

\subsubsection{Definitions} Now we state some important definitions for the AWGN-MAC with feedback. Please refer to \cite[Fig.~17.4]{elgamal} for an illustration of the setup of the problem.
\begin{definition}
\label{mac-def3}
An {\em $(n,M_{1n},M_{2n},\eps_n)$-code} for the AWGN-MAC with feedback and expected power constraints consists of two message sets $\calW_1 :=\{1,\ldots, {M}_{1n}\}$, $\calW_2=\{1,\ldots, {M}_{2n}\}$,   encoders $f_{1n}:\calW_1\times \mathbb{R}^{n-1}\to \mathbb{R}^n,f_{2n}:\calW_2\times \mathbb{R}^{n-1}\to \mathbb{R}^n$ and a decoder $\varphi_n: \mathbb{R}^n \to \calW_1\times\calW_2$ satisfying the power constraints
\begin{align}
\label{eq:eq69}
P_{\mathrm{avg}}^{(1)}:=\frac{1}{n}\sum_{k=1}^n \bbE[X_{1k}^2]&=\frac{1}{n}\sum_{k=1}^n \bbE[f_{1k}^2(W_1,Y^{k-1})]\leq P_1,\\
\label{eq:eq70}
P_{\mathrm{avg}}^{(2)}:=\frac{1}{n}\sum_{k=1}^n \bbE[X_{2k}^2]&=\frac{1}{n}\sum_{k=1}^n \bbE[f_{2k}^2(W_2,Y^{k-1})]\leq P_2,\end{align}
and the error probability constraint
\begin{align}
\Pr((\hat{W}_1,\hat{W}_2)\ne (W_1,W_2)) \leq \eps_n,
\end{align} where the messages $W_1$ and $W_2$ are uniformly distributed on  $\calW_1$ and $\calW_2$  respectively, and $(\hat{W}_1,\hat{W}_2)=\varphi_n(Y^n)$ is the decoded message pair.
\end{definition}
Note that we use $W$  in~\eqref{eqn:Wchannel} to denote the  Gaussian MAC and $W_1$ and $W_2$ to denote the random messages.
\begin{definition}
Let $\calC(\rho; P_1, P_2,\eps)$ be the set of all rate pairs $(R_1, R_2)\in\bbR_{+}^2$ such that
\begin{align}
R_1 &\le \rvC\left(\frac{P_1(1-\rho^2)}{1-\eps}\right)\\
R_2 &\le \rvC\left(\frac{P_2(1-\rho^2)}{1-\eps}\right)\\
R_1+R_2 &\le  \rvC\left(\frac{P_1+P_2+2\rho\sqrt{P_1P_2}}{1-\eps}\right).
\end{align}
The {\em information $\eps$-capacity region} $\overline{\mathcal{C}}_{\mathrm{fb}}^*(P_1,P_2,\eps)\subset\bbR_+^2$ is defined  to be the set
\begin{align}
\overline{\mathcal{C}}_{\mathrm{fb}}^*(P_1,P_2,\eps):=\bigcup_{0\leq \rho\leq 1} \calC(\rho; P_1, P_2,\eps).
\end{align}
The {\em information capacity region} is defined as
\begin{align}
\overline{\mathcal{C}}_{\mathrm{fb}}^*(P_1,P_2):=\bigcap_{\eps>0}\overline{\mathcal{C}}_{\mathrm{fb}}^*(P_1,P_2,\eps)=\lim_{\eps\to 0}\overline{\mathcal{C}}^*_{\mathrm{fb}}(P_1,P_2,\eps),
\end{align} where the limit exists because of the monotonicity of $\overline{\mathcal{C}}^*_{\mathrm{fb}}(P_1,P_2,\eps)$.
\end{definition}
\begin{definition}
\label{mac-def5}
A pair of non-negative numbers $(R_1,R_2)$ is {\em $\eps$-achievable} if there exists a sequence of $(n,M_{1n},M_{2n},P_1,P_2,\eps_n)$-feedback codes such that
\begin{align}
\liminf_{n\to\infty}\frac{1}{n}\ln M_{jn}\geq R_j, j=1,2, \hspace{2mm}\mbox{and}\hspace{2mm} \limsup_{n\to\infty}\eps_n \leq \eps.
\end{align}
The  {\em $\eps$-capacity region} $\mathcal{C}_{\mathrm{fb}}^*(P_1,P_2,\eps) \subset\mathbb{R}_{+}^2$ is defined to be the closure of the set of all $\eps$-achievable rate pairs $(R_1,R_2)$. The  {\em capacity region} is defined as
\begin{align}
\mathcal{C}^*_{\mathrm{fb}}(P_1,P_2):=\bigcap_{\eps>0}\mathcal{C}^*_{\mathrm{fb}}(P_1,P_2,\eps)=\lim_{\eps\to 0}\mathcal{C}^*_{\mathrm{fb}}(P_1,P_2,\eps),
\end{align} where the limit exists because of the monotonicity of $\mathcal{C}^*_{\mathrm{fb}}(P_1,P_2,\eps)$.
\end{definition}
\subsection{Main Results}\label{subsection32}
\begin{theorem}
\label{mac-thm4}
For any $\rho \in [0,1]$, there exists a  sequence of $(n,M_{1n},M_{2n},\eps)$-codes for the AWGN channel with feedback under the expected power constraint such that
\begin{align}
\ln M_{1n} &\geq n \rvC\left(\frac{P_1(1-\rho^2)}{1-\eps}\right)+O(\ln\ln n), \label{eqn:m_ach1} \\
\ln M_{2n} &\geq n \rvC \left(\frac{P_2(1-\rho^2)}{1-\eps}\right)+O(\ln\ln n),  \\
\ln (M_{1n} M_{2n})&\geq n \rvC\left(\frac{P_1+P_2+2\rho\sqrt{P_1P_2}}{1-\eps}\right)+O(\ln\ln n). \label{eqn:m_ach3}
\end{align}
Conversely, for every sequence of  $(n,M_{1n},M_{2n},\eps)$-codes for the AWGN channel with feedback under the expected power constraint, the following inequalities hold for some $\rho \in [0,1]$
\begin{align}
\ln M_{1n} &\leq n\rvC\left(\frac{P_1(1-\rho^2)}{1-\eps}\right)+O(n^{2/3}),  \label{eqn:m_con1} \\
\ln M_{2n} &\leq n \rvC\left(\frac{P_2(1-\rho^2)}{1-\eps}\right)+O(n^{2/3}), \\
\ln (M_{1n}M_{2n}) &\leq n \rvC\left(\frac{P_1+P_2+2\rho\sqrt{P_1P_2}}{1-\eps}\right)+O(n^{2/3}).\label{eqn:m_con3}
\end{align}
As a consequence, the  $\eps$-capacity region is equal to the information $\eps$-capacity region, i.e.,
\begin{align}
\mathcal{C}_{\mathrm{fb}}^*(P_1,P_2,\eps)=\overline{\mathcal{C}}_{\mathrm{fb}}^*(P_1,P_2,\eps),
\end{align}
and the  capacity region is also equal to the information  capacity region~\cite{Ozarow}, i.e.,
\begin{align}
\mathcal{C}_{\mathrm{fb}}^*(P_1,P_2)=\overline{\mathcal{C}}_{\mathrm{fb}}^*(P_1,P_2). \label{eqn:cap_regions}
\end{align}
\end{theorem}
\begin{IEEEproof}
The proof of Theorem \ref{mac-thm4} directly follows from achievability statement in  Proposition~\ref{mac-lem5}  in   Subsection~\ref{sec:ach_mac} and converse statement in  Proposition~\ref{mac-lem11} in Subsection~\ref{sec:conv_mac}. We note that the equality in \eqref{eqn:cap_regions} is exactly Ozarow's result~\cite{Ozarow} so Theorem \ref{mac-thm4} is a generalization of~\cite{Ozarow}.
\end{IEEEproof}

We remark that the $\eps$-capacity region  (for positive $\eps$) is indeed enlarged compared to the case when $\eps=0$, i.e., the strong converse again fails to hold as expected. Indeed the powers $P_j$ are replaced by $P_j / (1-\eps)$ similarly to the single-user case. However, the proofs are substantially more involved. The proof of the inner bound (achievability) modifies Ozarow's coding scheme~\cite{Ozarow} with a simple randomization argument so that the error probability is asymptotically bounded above by $\eps$ instead of being required to vanish.  The proof of the outer bound (converse) requires   non-trivial combinations of  ideas from the meta-converse~\cite{PPV10} (see Appendix \ref{app:info_spec}), information spectrum analysis~\cite{Alajaji95, ChenAlajaji95, Han10} (see Lemma \ref{mac-lem9}) and Ozarow's original weak converse proof~\cite{Ozarow}.

We also note that bounds on the scaling of the second-order terms are established---the second-order terms scale  as $O(\ln\ln n)$  for the achievability part (see \eqref{eqn:m_ach1}--\eqref{eqn:m_ach3}). In fact for the achievability part, a simple inspection of the proofs shows that the second-order terms are lower bounded by  $-\frac{1}{2}\ln\ln n+O(1) $ for the marginal rates and $-\ln\ln n+ O(1)$ for the sum rate (see \eqref{choiceM1}, \eqref{choiceM2} and \eqref{choiceM12} in the proof of Proposition \ref{mac-lem5}). The second-order terms are upper bounded by   $O(n^{2/3})$ for the converse part (see \eqref{eqn:m_con1}--\eqref{eqn:m_con3}). Tightening the orders of the bounds and finding the constants (second-order coding rate region) appear  to be challenging tasks, but may be achieved by leveraging ideas from the single-user case in Section~\ref{sec:awgn}, e.g., the conversion of a code with expected power constraints to a code with peak power constraints. We defer this to future work.

Similar to the single-user case, our achievability also holds under the maximal error formalism, because the encoders can always utilize the common randomness obtained from one use of the feedback to relabel the message pairs according to the uniform distribution.
\subsection{Proof of Achievability}\label{subsection33}  \label{sec:ach_mac}
We start with a preliminary lemma.
\begin{lemma}\label{mac-lem20}
Consider the quartic equations
\begin{align}
\sum_{j=0}^4 a_j z^j &=0 , \label{eqn:a_quartic} \quad\mbox{and}\\
\sum_{j=0}^4 b_j z^j &=0. \label{eqn:b_quartic}
\end{align}
where the coefficients $\{(a_j,b_j): j = 0,1,\ldots, 4\}$ satisfy
\begin{align}
|b_j-a_j| \leq \frac{d}{n}, \label{eqn:diff_coeffs}
\end{align} for some finite  constant  $d\ge 0$.  Let $z^*$ be a real  solution to \eqref{eqn:a_quartic} assuming a real solution exists. Then, for all $n$ large enough, there exists a real solution to  \eqref{eqn:b_quartic}, namely $z_n^*$, such that
\begin{equation}
|z_n^*-z^*| \le\frac{c}{n}
\end{equation}
for some  finite  constant $c\ge 0$.
\end{lemma}
\begin{IEEEproof}
The proof of   Lemma \ref{mac-lem20}  is provided  in Appendix  \ref{app:prf_continuity}.
\end{IEEEproof}
\begin{proposition}\label{mac-lem5}
There exists  a sequence of $(n,M_{1n},M_{2n},\eps)$-codes for the AWGN channel with feedback under the expected power constraint satisfying \eqref{eqn:m_ach1}--\eqref{eqn:m_ach3} for any $\rho \in [0,1]$.
As a result, the  $\eps$-capacity region for the AWGN-MAC with feedback under an expected power constraint contains the information $\eps$-capacity region, i.e.,
\begin{align}
\mathcal{C}_{\mathrm{fb}}^*(P_1,P_2,\eps)  \supset  \overline{\mathcal{C}}_{\mathrm{fb}}^*(P_1,P_2,\eps). \label{eqn:mac_ach_set}
\end{align}
\end{proposition}
\begin{IEEEproof}
Let $\rho^*$ be the largest real solution in $(0,1)$ of the following equation
\begin{align}
\label{eq91new}
&1+\frac{P_1}{1-\eps}+\frac{P_2}{1-\eps}+2x\sqrt{\frac{P_1}{1-\eps}}\sqrt{\frac{P_2}{1-\eps}}\nn\\*
&\qquad =\left[1+\frac{P_1(1-x^2)}{1-\eps}\right]\left[1+\frac{P_2(1-x^2)}{1-\eps}\right].
\end{align}
Then by Lemma~\ref{mac-lem20}, the equation
\begin{align}
\label{eq:eq88}
&1+\frac{P_1}{1-\eps+ 1/n}+\frac{P_2}{1-\eps+ 1/n}+2x\sqrt{\frac{P_1}{1-\eps+1/n}}\sqrt{\frac{P_2}{1-\eps+1/n}}\nn\\*
&\qquad =\left[1+\frac{P_1(1-x^2)}{1-\eps+ 1/n}\right]\left[1+\frac{P_2(1-x^2)}{1-\eps+1/n}\right]
\end{align}
has a real solution $\rho_n^*$ such that $|\rho_n^*-\rho^*| \leq  {c(P_1,P_2,\eps)}/{n}$ for $n$ large enough. This is because \eqref{eq91new} and \eqref{eq:eq88} are quartic equations and each of their coefficients differ by no more than a constant of $1/n$.  Note that since $\rho^* \in (0,1)$, we also have $\rho_n^* \in (0,1)$ for $n$ large enough.

Similarly to the standard achievability proof  (e.g., \cite{Ozarow} and \cite[Sec. 17.2.4]{elgamal}) for the vanishing error probability formalism, for each fixed $n$ we will first show that \eqref{eqn:m_ach1}--\eqref{eqn:m_ach3} hold for $\rho = \rho_n^*$, where  $\rho_n^*  \in (0,1)$ is the solution of~\eqref{eq:eq88}.

As usual $|\mathcal{W}_1|=M_{1n},$ and $|\mathcal{W}_2|=M_{2n}$. At a high-level, we combine Ozarow's coding scheme~\cite{Ozarow} and power control ideas from Yang {\em et al.}~\cite{yang15} with some modifications. More specifically, for each pair $(w_1,w_2) \in \mathcal{W}_1\times \mathcal{W}_2$, the coding scheme is as follows:

\subsubsection{Encoding}
\begin{itemize}
\item In the first transmission, the two transmitters send zero symbols, i.e., $X_{11}=0, X_{21}=0$. They receive the first feedback signals which are equal to the first channel noise symbol via the feedback links, i.e., $Y_1=Z_1$.
\item For the next $n$ transmissions, we first  define  the rates
\begin{align}
R_{1n} &:=\frac{1}{n}{\ln M_{1n}} , \label{eqn:defR1}\\
R_{2n} &:=\frac{1}{n} {\ln M_{2n}}. \label{eqn:defR2}
\end{align}
As in    Ozarow's paper~\cite[pp.\ 625] {Ozarow}, at time $k=1,2$ the receiver adds to his received outputs a   random  variable $W\sim \calN(0,\sigma_W^2)$. The ``degraded" outputs are fed back to the transmitters and used at both ends to form estimates. Define a sequence $\rho_k, k=2,3,...,n$ as~\cite[Equation (4)]{Ozarow}. Then, if we choose $\sigma_W^2$ such that
\begin{align}
\label{eq99new}
\frac{\bbE\left[\frac{Z_1+W}{\sqrt{12P_1/(1-\eps+1/n)}}\frac{Z_2+W}{\sqrt{12P_2/(1-\eps+1/n)}}\right]}{\sqrt{\left(\frac{1+\sigma_W^2}{12P_1/(1-\eps+1/n)}\right)\left(\frac{1+\sigma_W^2} {12P_2(1-\eps+1/n)}\right)}}=\rho_n^*,
\end{align} we will have $\rho_2=\rho_n^*$. Note that the equation~\eqref{eq99new} is equivalent to
\begin{align}
\frac{\sigma_W^2}{1+\sigma_W^2}=\rho_n^*,
\end{align} or
\begin{align}
\sigma_W^2=\frac{\rho_n^*}{1-\rho_n^*}.
\end{align}

Since $\rho_2=\rho_n^*$, similarly to the argument leading to~\cite[Equation (11)]{Ozarow} we have $\rho_k=(-1)^k \rho_n^*$ for all $k=2,3\ldots,n$. Therefore, an upper bound on the average error probability associated to the  Ozarow scheme~\cite[Equation (13)]{Ozarow}  is
\begin{align}
\kappa_n:& = 2\rmQ\left[\frac{1}{2\sqrt{v_{1n}}\left(1+\frac{P_1[1-(\rho_n^*)^2]}{1-\eps+1/n}\right)}\exp\left[n\left(\rvC\left(1+\frac{P_1[1-(\rho_n^*)^2]}{1-\eps+1/n}\right)-R_{1n}\right)\right]\right]\nonumber\\
&\qquad +2\rmQ\left[\frac{1}{2\sqrt{v_{2n}}\left(1+\frac{P_2[1-(\rho_n^*)^2]}{1-\eps+1/n}\right)}\exp\left[n\left(\rvC\left(1+\frac{P_2[1-(\rho_n^*)^2]}{1-\eps+1/n}\right)-R_{2n}\right)\right]\right]\label{defEpsN0}
\end{align}
where $v_{1n}=(1-\eps+1/n)/(12P_1), v_{2n}=(1-\eps+1/n)/(12P_2)$, the complementary Gaussian cdf $\rmQ(u):=1-\Phi(u)$. Now using the Chernoff bound on $\rmQ(\cdot) $ to upper bound \eqref{defEpsN0} (similarly to \eqref{eqn:cher}), we obtain
\begin{align}
\kappa_n&\leq \exp\left[\frac{-1}{8v_{1n}\left(1+\frac{P_1[1-(\rho_n^*)^2]}{1-\eps+1/n}\right)^2}\exp\left[2n\left(\rvC\left(1+\frac{P_1[1-(\rho_n^*)^2]}{1-\eps+1/n}\right)-R_{1n}\right)\right]\right] \nonumber\\
&\qquad +\exp\left[\frac{-1}{8v_{2n}\left(1+\frac{P_2[1-(\rho_n^*)^2]}{1-\eps+1/n}\right)^2}\exp\left[2n\left(\rvC\left(1+\frac{P_2[1-(\rho_n^*)^2]}{1-\eps+1/n}\right)-R_{2n}\right)\right]\right].
\label{defEpsN}
\end{align}

In the following, we will design the parameters of the code so that $\kappa_n\to 0$. Henceforth, we assume that $n$ is sufficiently  large so that $\kappa_n<\eps$ and $\frac{1}{n}<\eps$ (so subsequent  expressions make sense).

 Now we also adopt the following strategy:
\begin{itemize}
\item If
\begin{equation}
Y_1 \le  \Phi^{-1}\left(\frac{\eps-\kappa_n}{1-\kappa_n}\right), \label{eqn:zero_syms}
\end{equation}
 then both the encoders send zero symbols for all the $n$ transmission, i.e., $\{X_{1k}\}_{k=2}^{n+1}=  \{X_{2k}\}_{k=2}^{n+1}=(0,\ldots ,0)$.
\item If
\begin{equation}
Y_1  > \Phi^{-1}\left(\frac{\eps-\kappa_n}{1-\kappa_n}\right), \label{eqn:non_zero_syms}
\end{equation}
 then encoder $j$ for $j\in \{1,2\}$ sends the next $n$ transmission symbols following Ozarow's coding scheme with expected power constraint ${P_j}/ {(1-\eps+ 1/n)}$.
\end{itemize}
\end{itemize}

\subsubsection{Decoding}
\begin{itemize}
\item For the first received signal symbols $Y_1=Z_1$, the receiver feed backs this signal to transmitters via the feedback links.
\item For the next $n$ received signals, $Y_k=X_{1k}+X_{2k} +Z_k, k=2,3,...,n+1$, the receiver feed backs the received signals to the transmitters via feedback links and performs decoding as Ozarow's decoding algorithm.
\end{itemize}

Here, we remark that the first noise variable $Z_1=Y_1$ is used as a ``biased coin flip'' to either transmit (if \eqref{eqn:non_zero_syms}  is true) or not (if instead \eqref{eqn:zero_syms} is true).  This ensures that all encoders $f_k$ are deterministic. Furthermore, as we shall see in the error probability analysis to follow, the choices of  various parameters ensure  that power constraints and error probability bound are simultaneously satisfied.

\subsubsection{Error Probability Analysis}

First define the event  $\mathcal{E}_{\eps,\kappa_n}:=\big\{Y_1\le  \Phi^{-1}(\frac{\eps-\kappa_n}{1-\kappa_n}) \big\}$, i.e., this event implies both encoders transmit zero symbols over times $k = 2,3,\ldots, n+1$ according to \eqref{eqn:zero_syms}. Since $Y_1$ is standard Gaussian,  we have $\Pr(\mathcal{E}_{\eps,\kappa_n})=\frac{\eps-\kappa_n}{1-\kappa_n}$ and $\Pr( \calE_{\eps,\kappa_n}^c ) = \frac{1-\eps}{1-\kappa_n}$. By Ozarow's analysis~\cite{Ozarow}, we have
\begin{align}
\label{eq:eq87}
\Pr\left(\{\hat{W}_1 \neq W_1\} \cup \{\hat{W}_2\neq W_2\}\,\Big|\,  \mathcal{E}_{\eps,\kappa_n}^c\right) \leq \kappa_n.
\end{align}
It follows that
\begin{align}
&\Pr\left(\{\hat{W}_1 \neq W_1 \} \cup \{\hat{W}_2\neq W_2\}\right) \\
&\leq \Pr\left(\{\hat{W}_1 \neq W_1\} \cup \{\hat{W}_2\neq W_2\}\,\Big|\,  \mathcal{E}_{\eps,\kappa_n}^c\right) \Pr\left(\mathcal{E}_{\eps,\kappa_n}^c\right) + \Pr\left(\mathcal{E}_{\eps,\kappa_n}\right),\label{eqn:condE}\\
\label{eq:eq100a}
&=\kappa_n \left(\frac{1-\eps}{1-\kappa_n}\right) + \frac{\eps-\kappa_n}{1-\kappa_n }\\
& = \eps
\end{align}
where \eqref{eq:eq100a} follows from the bounds on the probability of $\calE_{\eps,\kappa_n}$ and its complement, and the fact that the average error probability of the Ozarow scheme is upper bounded by $\kappa_n$ in \eqref{defEpsN}. Hence, the average error probability of the coding scheme is upper bounded by $\eps$.

Following~\eqref{defEpsN}, we can show by using Taylor expansion that
\begin{align}
&\rvC\left(\frac{P_1[1-(\rho_n^*)^2]}{1-\eps+1/n}\right)= \rvC\left(\frac{P_1[1-(\rho_n^*)^2]}{1-\eps}\right)+O\left(\frac{1}{n}\right),\\
&\rvC\left(\frac{P_2[1-(\rho_n^*)^2]}{1-\eps+1/n}\right)= \rvC\left(\frac{P_2[1-(\rho_n^*)^2]}{1-\eps}\right)+O\left(\frac{1}{n}\right), \end{align}
and
\begin{align}
&\rvC\left(\frac{P_1}{1-\eps+1/n}+\frac{P_2}{1-\eps+1/n}+ 2\rho_n^* \sqrt{\left(\frac{P_1}{1-\eps+1/n}\right)\left(\frac{P_2}{1-\eps+1/n}\right)}\right) \nonumber \\
&=\rvC\left(\frac{P_1}{1-\eps}+\frac{P_2}{1-\eps}+ 2\rho_n^* \sqrt{\left(\frac{P_1}{1-\eps}\right)\left(\frac{P_2}{1-\eps}\right)}\right)+O\left(\frac{1}{n}\right).
\end{align}
Now we choose $\vartheta_n=O(1)$ (to be determined later) and
\begin{align}
\ln M_{1n} &=n \rvC\left(\frac{P_1[1-(\rho_n^*)^2]}{1-\eps+1/n}\right)-\frac{1}{2}\ln(\vartheta_n\ln n),\label{choiceM1} \\
\ln M_{2n} &=n \rvC\left(\frac{P_2[1-(\rho_n^*)^2]}{1-\eps+1/n}\right)- \frac{1}{2} \ln(\vartheta_n\ln n), \label{choiceM2}
\end{align}
where $\rho_n^*$ is the solution of~\eqref{eq:eq88}. Using \eqref{eq:eq88}, \eqref{choiceM1} and~\eqref{choiceM2}, we have
%
\begin{align}
\ln (M_{1n}M_{2n})&=\ln M_{1n} + \ln M_{2n}\\
&=n \rvC\left(\frac{P_1[1-(\rho_n^*)^2]}{1-\eps+1/n}\right)+ n \rvC\left(\frac{P_2[1-(\rho_n^*)^2]}{1-\eps+1/n}\right)-\ln(\vartheta_n\ln n)\\
&=n\rvC\left(\frac{P_1+P_2 +2 \rho_n^*\sqrt{P_1 P_2}}{1-\eps+1/n}\right)-\ln(\vartheta_n\ln n). \label{choiceM12}
\end{align}
Since $|\rho_n^*-\rho^*|\leq c(P_1,P_2,\eps)/n$ for $n$ large enough, by Taylor expansions,  we have
\begin{align}
\ln M_{1n} &=n \rvC\left(\frac{P_1[1-(\rho^*)^2]}{1-\eps}\right)-\frac{1}{2}\ln \ln n+O(1),\label{choiceM1new} \\
\ln M_{2n} &=n \rvC\left(\frac{P_2[1-(\rho^*)^2]}{1-\eps}\right)-\frac{1}{2}\ln \ln n+O(1),\label{choiceM2new} \\
\ln (M_{1n}M_{2n})&=n\rvC\left(\frac{P_1+P_2 +2 \rho^*\sqrt{P_1 P_2}}{1-\eps}\right)-\ln \ln n +O(1). \label{choiceM12new}
\end{align}

In addition, by our choices of the various parameters, we note that (cf.\ the definition of the rates in \eqref{eqn:defR1}--\eqref{eqn:defR2})
\begin{align}
\rvC\left(\frac{P_1[1-(\rho_n^*)^2]}{1-\eps+1/n}\right)-R_{1n} &=  \frac{\ln(\vartheta_n\ln n)}{2n}, \label{tempEq1}\\
\rvC\left(\frac{P_2[1-(\rho_n^*)^2]}{1-\eps+1/n}\right)-R_{2n} &=  \frac{\ln(\vartheta_n \ln n)}{2n} ,\label{tempEq2}
\end{align}
so by \eqref{defEpsN} we have
\begin{align}
\kappa_n &\leq \exp\left[\frac{-\vartheta_n \ln n}{8v_{1n}\left(1+\frac{P_1[1-(\rho_n^*)^2]}{1-\eps+1/n}\right)^2}  \right]  +\exp\left[\frac{-\vartheta_n \ln n}{8v_{2n}\left(1+\frac{P_2[1-(\rho_n^*)^2]}{1-\eps+1/n}\right)^2} \right].
\end{align}
Let
\begin{align}
\vartheta_n:= 2\max\left\{8v_{1n}\left(1+\frac{P_1[1-(\rho_n^*)^2]}{1-\eps+1/n}\right)^2, 8v_{2n}\left(1+\frac{P_2[1-(\rho_n^*)^2]}{1-\eps+1/n}\right)^2\right\}.
\end{align} Note that this parameter behaves as $\vartheta_n=O(1)$. We then obtain
\begin{align}
\kappa_n &\leq 2 \exp\left(-2 \ln n \right)\\
&=\frac{2}{n^2}.
\end{align}
It follows that
\begin{align}
\kappa_n\Big(1-\eps+\frac{1}{n}\Big)\leq \frac{1}{n}\label{eqn:kappa_bd} .
\end{align}

With regard to the input power constraints, we may conclude from~\eqref{eqn:kappa_bd} that the average transmitted powers at encoders $1$ and $2$ satisfy
\begin{align}
P_{\mathrm{avg}}^{(1)} &\leq  \frac{1-\eps}{1-\kappa_n} \left(\frac{P_1}{1-\eps+1/n}\right)\\
&\leq P_1,\\
P_{\mathrm{avg}}^{(2)}&\leq \frac{1-\eps}{1-\kappa_n}\left(\frac{P_2}{1-\eps+1/n}\right)\\
&\leq P_2.
\end{align}
Hence, the input power constraints are satisfied.
%
Consequently, there exists an $(n,M_{1n},M_{2n},\eps)$-code for the AWGN-MAC   with feedback satisfying \eqref{eqn:m_ach1}--\eqref{eqn:m_ach3} for $\rho = \rho^*$ for large enough~$n$. Following standard arguments (e.g., \cite{Ozarow} and \cite[Sec. 17.2.4]{elgamal}) which involve partitioning one of the message sets and using successive cancellation decoding, we can conclude that for each $\rho\in[0, 1]$, there exists an $(n,M_{1n},M_{2n},\eps)$-code for the AWGN-MAC with feedback satisfying \eqref{eqn:m_ach1}--\eqref{eqn:m_ach3} for large enough~$n$. This means that the  $\eps$-capacity region satisfies the following inclusion
\begin{align}
\mathcal{C}_{\mathrm{fb}}^*(P_1,P_2,\eps)\supset \overline{\mathcal{C}}_{\mathrm{fb}}^*(P_1,P_2,\eps),
\label{eqn:eq83}
\end{align}
completing the proof of Proposition \ref{mac-lem5} (i.e., the achievability part of Theorem \ref{mac-thm4}).
\end{IEEEproof}
\subsection{Proof of Converse } \label{sec:conv_mac}

\paragraph*{Outline}
To establish an outer bound for the $\eps$-capacity region, we use Lemmas \ref{mac-lem6} to   \ref{mac-lem10} to establish Proposition \ref{mac-lem11}. In particular, Lemma \ref{mac-lem6} is an important ``single-letterization'' lemma that allows us to amalgamate all the different correlation parameters $\rho_k$ (the correlation coefficients between the input symbols $X_{1k}$ and $X_{2k}$) and to introduce a {\em single} parameter $\rho$ whose magnitude does not exceed $1$.  The idea behind Lemma \ref{mac-lem6} is partly inspired by the weak converse proof for the AWGN-MAC with feedback by Ozarow~\cite{Ozarow}.  Lemma \ref{mac-lem7} allows us to bound the probabilities of certain atypical events given power constraints on the inputs.  Lemma \ref{mac-lem8} provides   computations of the moments of certain important statistics. In particular, it shows that the second moments of certain important information density random variables scales as $O(n)$. Lemma \ref{mac-lem9}  provides important information spectrum-type  upper bounds on the maximum number of codewords transmissible with $\eps$ error. Lemma \ref{mac-lem10}  bounds the probabilities within the information spectrum-type  upper bound by using the moment calculations in Lemma  \ref{mac-lem8}. Finally Proposition \ref{mac-lem11} puts the preceding calculations together to establish the outer bound on the $\eps$-capacity region.

\begin{lemma}
\label{mac-lem6}
Consider a feedback code  for the AWGN-MAC of length $n$ with encoders $\{(f_{1k}, f_{2k})\}_{k=1}^n$ that yields input symbols $X_{1k}=f_{1k}(W_1,Y^{k-1})$ and $X_{2k}=f_{2k}(W_2,Y^{k-1})$ which have second moments
\begin{equation}
P_{1k}:=\bbE[X_{1k}^2] ,\quad\mbox{and}\quad P_{2k}:= \bbE[X_{2k}^2].
\end{equation}
 Define
\begin{align}
\label{mac-key-point}
\rho:=\frac{\sum_{k=1}^n \rho_k\sqrt{ P_{1k}P_{2k}}}{n\sqrt{P_1P_2}},
\end{align} where the $\rho_k$'s are correlation coefficients defined by
\begin{align}
\label{mac-rhok}
\rho_k:=\frac{\bbE[X_{1k} X_{2k}]}{\sqrt{P_{1k}P_{2k}}}, \qquad k=1,2,\ldots ,n.
\end{align}
Then,
\begin{align}
\label{eq:eq123}
|\rho| &\le 1.
\end{align}
Furthermore,
\begin{align}
\label{eq:eq124}
\sum_{k=1}^n P_{1k}(1-\rho_k^2) &\leq nP_1(1-\rho^2),\\
\label{eq:eq125}
\sum_{k=1}^n P_{2k}(1-\rho_k^2) &\leq nP_2(1-\rho^2),\\
\label{eq:eq126}
\sum_{k=1}^n P_{1k}+ P_{2k}+2\rho_k \sqrt{P_{1k}P_{2k}} &\leq n(P_1+P_2+2\rho\sqrt{P_1P_2}).
\end{align}
\end{lemma}
\begin{IEEEproof}
Observe that
\begin{align}
|\rho|&=\left|\frac{\sum_{k=1}^n\rho_k \sqrt{P_{1k}P_{2k}}}{n\sqrt{P_1P_2}}\right|\\
&\stackrel{(a)}{\leq}  \frac{1}{n} \frac{\sum_{k=1}^n \sqrt{P_{1k}P_{2k}}}{\sqrt{P_1P_2}}\\
&=\frac{1}{n}\sum_{k=1}^n \sqrt{\frac{P_{1k}P_{2k}}{P_1P_2}} \\
&\stackrel{(b)}{\leq} \frac{1}{2n}\sum_{k=1}^n \left[\frac{P_{1k}}{P_1}+\frac{P_{2k}}{P_2}\right]\\
&=\sum_{k=1}^n \frac{1}{2n} \frac{P_{1k}}{P_1}+\frac{1}{2n}\sum_{k=1}^n \frac{P_{2k}}{P_2}\\
&\stackrel{(c)}{\leq} \frac{1}{2n}\cdot\frac{nP_1}{P_1}+ \frac{1}{2n}\cdot\frac{nP_2}{P_2}\\
\label{eq:eq133}
&=1.
\end{align}
Here, (a) follows from the fact that $|\rho_k|\le 1$ (by the Cauchy-Schwarz inequality),  (b) follows from the arithmetic mean-geometric mean inequality ($\sqrt{ab}\le\frac{a+b}{2}$), and (c) follows from the input power constraints~\eqref{eq:eq69} and~\eqref{eq:eq70}. This verifies \eqref{eq:eq123}.

In addition, from the definition of $\rho$ in~\eqref{mac-key-point} we also have
\begin{align}
\sum_{k=1}^n P_{1k}+ P_{2k}+2\rho_k \sqrt{P_{1k}P_{2k}} &\leq nP_1 + nP_2 + 2\sum_{k=1}^n \rho_k \sqrt{P_{1k}P_{2k}}\\
\label{eq:eq135}
&= n(P_1+P_2+2\rho\sqrt{P_1P_2}),
\end{align}
verifying \eqref{eq:eq126}.
Moreover, we see that
\begin{align}
(n\rho\sqrt{P_1P_2})^2 &= \left(\sum_{k=1}^n \rho_k \sqrt{P_{1k}P_{2k}}\right)^2\\
&\stackrel{(a)}{\leq} \left(\sum_{k=1}^n P_{1k} \rho_k^2\right)\left(\sum_{k=1}^n P_{2k}\right) \\
&\stackrel{(b)}{\leq} \left(\sum_{k=1}^n P_{1k} \rho_k^2\right) nP_2
\end{align}
Here, (a) follows from the Cauchy--Schwarz inequality, (b) follows from the input power constraint~\eqref{eq:eq70}. Therefore, we obtain
\begin{align}
\label{eq:eq139}
\sum_{k=1}^n P_{1k} \rho_k^2 \geq nP_1\rho^2.
\end{align}
It follows that
\begin{align}
\sum_{k=1}^n P_{1k}(1-\rho_k^2)&=\sum_{k=1}^n P_{1k} -\sum_{k=1}^n P_{1k}\rho_k^2\\
&\stackrel{(a)}{\leq} nP_1 - \sum_{k=1}^n P_{1k}\rho_k^2\\
&\stackrel{(b)}{\leq} nP_1 - nP_1\rho^2\\
\label{eq:eq142}
&=nP_1(1-\rho^2),
\end{align}
where (a) is due to the input power constraint~\eqref{eq:eq69} and (b) follows from~\eqref{eq:eq139}.  This verifies \eqref{eq:eq124}. Using the same arguments, we also show that
\begin{align}
\label{eq:eq144}
\sum_{k=1}^n P_{2k}(1-\rho_k^2)\leq nP_2(1-\rho^2).
\end{align}
This completes the proof.
\end{IEEEproof}
\begin{lemma}
\label{mac-lem7}
Assume that all parameters are defined as Lemma~\ref{mac-lem6}. Define the events
\begin{align}
\mathcal{U}_{1n}&:=\left\{\sum_{k=1}^n\left(X_{1k}-X_{2k}\rho_k\sqrt{\frac{P_{1k}}{P_{2k}}}\right)^2-n P_1T(1-\rho^2)< 0\right\}, \\
\mathcal{U}_{2n}&:=\left\{\sum_{k=1}^n\left(X_{2k}-X_{1k}\rho_k\sqrt{\frac{P_{2k}}{P_{1k}}}\right)^2-n P_2T(1-\rho^2)< 0\right\},  \\
\mathcal{U}_{3n}&:=\left\{\sum_{k=1}^n(X_{1k}+X_{2k})^2-n(P_1 T+P_2T+2\rho\sqrt{P_1TP_2T})< 0\right\},
\end{align} for some positive real number $T$. Then, the following inequalities hold:
\begin{align}
\label{mac-key-point2}
\Pr(\mathcal{U}_{jn}^c)\leq \frac{1}{T},\qquad  j=1,2,3.
\end{align}
\end{lemma}
\begin{IEEEproof}
 Observe that
\begin{align}
\Pr(\mathcal{U}_{1n}^c) &\stackrel{(a)}{\leq}\frac{\bbE\left[\sum_{k=1}^n\left(X_{1k}-X_{2k}\rho_k\sqrt{\frac{P_{1k}}{P_{2k}}}\right)^2\right]}{n P_1T(1-\rho^2)}\\
&=\frac{\sum_{k=1}^n P_{1k}(1-\rho_k^2)}{n P_1T(1-\rho^2)}\\
&\stackrel{(b)}{\leq} \frac{nP_1(1-\rho^2)}{nP_1T(1-\rho^2)}\\
&=\frac{1}{T},
\end{align} where (a) follows from Markov's inequality \cite[Prop.~1.1]{TanBook} and (b) follows from~\eqref{eq:eq124}. Similarly, we also have
\begin{align}
\label{eq:eq140}
\Pr(\mathcal{U}_{2n}^c) \leq  \frac{1}{T}.
\end{align}
In addition, we see that
\begin{align}
\label{eq:eq141}
\Pr(\mathcal{U}_{3n}^c)&\stackrel{(a)}{\leq} \frac{\sum_{k=1}^n \bbE[(X_{1k}+X_{2k})^2]}{nT(P_1+P_2+2\rho\sqrt{P_1P_2})} \\
&=\frac{\sum_{k=1}^n P_{1k}+P_{2k}+2\rho_k \sqrt{P_{1k}P_{2k}}}{nT(P_1+P_2+2\rho\sqrt{P_1P_2})}\\
&\stackrel{(b)}{\leq} \frac{nP_1 + nP_2 + 2\sum_{k=1}^n \rho_k \sqrt{P_{1k}P_{2k}}}{nT(P_1+P_2+2\rho\sqrt{P_1P_2})}\\
&\stackrel{(c)}{=} \frac{n(P_1+P_2+2\rho\sqrt{P_1P_2})}{nT(P_1+P_2+2\rho\sqrt{P_1P_2})}\\
&=\frac{1}{T}.
\end{align} where (a) follows from   Markov's inequality \cite[Prop.~1.1]{TanBook}, (b) follows from the input power constraints~\eqref{eq:eq69},~\eqref{eq:eq70}, and (c) follows from~\eqref{eq:eq126}.
\end{IEEEproof}
\begin{lemma}
\label{mac-lem8}
Consider the  parameters as defined as Lemma~\ref{mac-lem6}. Define the random variables.
\begin{align}
V_{1n} &:=-P_1T(1-\rho)^2\sum_{k=1}^n Z_k^2 + 2\sum_{k=1}^n Z_k\left(X_{1k}-\rho_k X_{2k}\sqrt{\frac{P_{1k}}{P_{2k}}}\right)+nP_1T(1-\rho^2),\\
V_{2n} &:=-P_2T(1-\rho)^2\sum_{k=1}^n Z_k^2 + 2\sum_{k=1}^n Z_k\left(X_{2k}-\rho_k X_{1k}\sqrt{\frac{P_{2k}}{P_{1k}}}\right)+nP_2T(1-\rho^2),\\
V_{3n} &:= 2\sum_{k=1}^n (X_{1k}+X_{2k}) Z_k -(P_1 T+P_2 T+2\rho T\sqrt{P_1P_2})\sum_{k=1}^n Z_k^2+n(P_1 T+P_2 T+2\rho T\sqrt{P_1P_2}).
\end{align} Then  their moments are given as
\begin{align}
\bbE[V_{jn}]&=0,\qquad  j=1,2,3,\\
\label{eq:eq163}
\bbE[V_{1n}^2]&=2nP_1^2 T^2(1-\rho^2)^2 + 4 \sum_{k=1}^n P_{1k}(1-\rho_k^2),\\
\label{eq:eq164}
\bbE[V_{2n}^2]&=2nP_2^2 T^2(1-\rho^2)^2 + 4 \sum_{k=1}^n P_{2k}(1-\rho_k^2),\\
\label{eq:eq165}
\bbE[V_{3n}^2]&=4\sum_{k=1}^n (P_{1k}+P_{2k}+2\rho_k\sqrt{P_{1k}P_{2k}})+2n(P_1T+P_2T+2\rho T\sqrt{P_1P_2})^2.
\end{align}
\end{lemma}
\begin{IEEEproof}
Since the proof is straightforward but tedious, we defer it to Appendix~\ref{app:prf_moments}.

Now we state an information spectrum-type lemma that is similar to \cite[Lemma 4]{Han98} and \cite[Proposition 1]{ChenAlajaji95} but adapted to suit the needs of the problem at hand.
\begin{lemma}
\label{mac-lem9}
Consider any length-$n$ code for the stationary memoryless MAC $P_{Y^n |X_1^nX_2^n}(y^n |x_1^n,x_2^n)=W^n (y^n |x_1^n,x_2^n)=\prod_{k=1}^n W(y_k| x_{1k}, x_{2k})$ with feedback (cf.~Definition~\ref{mac-def3}). This induces a code distribution $P_{W_1 W_2 X_1^n X_2^n} \times W^n$ defined in terms of the encoders $\{ f_{1k}, f_{2k} \}_{k=1}^n$. Then for any positive real numbers $\gamma_{1n}, \gamma_{2n}, \gamma_{3n}$ and any collection of (output) distributions $\{(Q_{Y_k|X_{1k}}, Q_{Y_k| {X_{2k}}}, Q_{Y_k})\}_{ k=1}^n$, the following bounds hold:
\begin{align}
\label{eq:eq194}
\ln M_{1n} \leq& \ln \gamma_{1n}-\ln^{+}\left(1-\eps-\Pr\left[\sum_{k=1}^n \ln \frac{W(Y_k|X_{1k} X_{2k})}{Q_{Y_k|X_{2k}}(Y_k|X_{2k})} \geq \ln\gamma_{1n}\right] \right),\\
\label{eq:eq195}
\ln M_{2n} \leq& \ln \gamma_{2n}-\ln^{+}\left(1-\eps-\Pr\left[\sum_{k=1}^n \ln \frac{W(Y_k|X_{1k} X_{2k})}{Q_{Y_k|X_{1k}}(Y_k|X_{1k})} \geq \ln\gamma_{2n}\right] \right),\\
\label{eq:eq196}
\ln \left(M_{1n}M_{2n}\right) \leq& \ln \gamma_{3n}-\ln^{+}\left(1-\eps-\Pr\left[\sum_{k=1}^n \ln \frac{W(Y_k|X_{1k} X_{2k})}{Q_{Y_k}(Y_k)} \geq \ln\gamma_{3n}\right] \right).
\end{align}
\end{lemma}
\begin{IEEEproof}
See  Appendix \ref{app:info_spec}.
\end{IEEEproof}
\begin{lemma}
\label{mac-lem10}
Given a positive real number $1/(1-\eps) \le T\le  2/(1-\eps)$  and $\rho$ as defined in Lemma~\ref{mac-lem6}, the following inequalities hold for some choice of (output) distributions $\{(Q_{Y_k|X_{1k}}, Q_{Y_k| {X_{2k}}}, Q_{Y_k})\}_{k=1}^n$:
\begin{align}
\Pr\left[\sum_{k=1}^n \ln \frac{W(Y_k|X_{1k} X_{2k})}{Q_{Y_k|X_{2k}}(Y_k|X_{2k})} \geq \ln\gamma_{1n}\right]&\leq \frac{1}{T}+ O(n^{-1/3}) \label{eqn:O1}\\
\Pr\left[\sum_{k=1}^n \ln \frac{W(Y_k|X_{1k} X_{2k})}{Q_{Y_k|X_{1k}}(Y_k|X_{1k})} \geq \ln\gamma_{1n}\right]&\leq \frac{1}{T}+ O(n^{-1/3}) \\
\Pr\left[\sum_{k=1}^n \ln \frac{W(Y_k|X_{1k} X_{2k})}{Q_{Y_k}(Y_k)}\geq \ln\gamma_{3n}\right] &\leq \frac{1}{T}+O(n^{-1/3}),\label{eqn:O3}
\end{align} where $\gamma_{1n}, \gamma_{2n}$ and $\gamma_{3n}$ are defined as
\begin{align}
\ln\gamma_{1n} &= \frac{n}{2} \ln\left[1+P_1T(1-\rho^2)\right]+n^{2/3},\\
\ln\gamma_{2n} &= \frac{n}{2} \ln\left[1+P_2T(1-\rho^2)\right]+n^{2/3},\\
\ln \gamma_{3n}&=\frac{n}{2}\ln\left[1+P_1T+P_2T+2\rho T\sqrt{P_1P_2}\right]+ n^{2/3}.
\end{align}
The implied constants in the $O(\cdot)$ notation in~\eqref{eqn:O1}--\eqref{eqn:O3} depend only on $P_1, P_2$ and $\eps$.
\end{lemma}

\begin{IEEEproof}
Firstly, we choose the auxiliary conditional output distributions in the statement of Lemma \ref{mac-lem9} to be
\begin{align}
Q_{Y_k|X_{2k}}(y_k|x_{2k})&: =\mathcal{N}\left(y_k;0, x_{2k}\Big(1+\rho_k \sqrt{\frac{P_{1k}}{P_{2k}}}\Big), 1+P_1 T(1-\rho^2)\right),\\
Q_{Y_k|X_{1k}}(y_k|x_{1k})&:=\mathcal{N}\left(y_k;0, x_{1k}\Big(1+\rho_k \sqrt{\frac{P_{2k}}{P_{1k}}}\Big), 1+P_2 T(1-\rho^2)\right),\\
Q_{Y_k}(y_k)&:=\mathcal{N} \left(y_k;0, 1+P_1 T+P_2 T+2\rho T\sqrt{P_1 P_2} \right),
\end{align} for all $k=1,2,\ldots ,n$. Here, $\rho_k$ and $\rho$ were defined in \eqref{mac-rhok} and \eqref{mac-key-point} respectively and we also use  the notation $\mathcal{N}(y;\mu,\sigma):=(2\pi\sigma^2)^{-1/2}\exp\big(-{(y-\mu)^2}/{(2\sigma)}\big)$ for the normal distribution.

Observe that
\begin{align}
&\sum_{k=1}^n \ln \frac{W(Y_k|X_{1k} X_{2k})}{Q_{Y_k|X_{2k}}(Y_k|X_{2k})}\nn\\*
&=\frac{n}{2} \ln\left[1+P_1T(1-\rho^2)\right]
+\frac{-[1+P_1T(1-\rho^2)]\sum_{k=1}^n Z_k^2 +\sum_{k=1}^n \left[Y_k-X_{2k}\left(1+\rho_k\sqrt{\frac{P_{1k}}{P_{2k}}}\right)\right]^2}{2[1+P_1T(1-\rho^2)]},\\
&=\frac{n}{2} \ln\left[1+P_1T(1-\rho^2)\right] +\frac{\left[\sum_{k=1}^n\left(X_{1k}-X_{2k}\rho_k\sqrt{\frac{P_{1k}}{P_{2k}}}\right)^2-n P_1T(1-\rho^2)\right]}{2(1+P_1T(1-\rho^2))} \nonumber \\
&\qquad +\frac{\left[-P_1T(1-\rho)^2\sum_{k=1}^n Z_k^2 + 2\sum_{k=1}^n Z_k\left(X_{1k}-\rho_k X_{2k}\sqrt{\frac{P_{1k}}{P_{2k}}}\right)+nP_1T(1-\rho^2)\right]}{2(1+P_1T(1-\rho^2))}.
\end{align}
Similarly, we also have
\begin{align}
&\sum_{k=1}^n \ln \frac{W(Y_k|X_{1k} X_{2k})}{Q_{Y_k|X_{1k}}(Y_k|X_{1k})}\nn\\*
&=\frac{n}{2} \ln \left[1+P_2T(1-\rho^2)\right]+\frac{\left[\sum_{k=1}^n\left(X_{2k}-X_{1k}\rho_k\sqrt{\frac{P_{2k}}{P_{1k}}}\right)^2-n P_2T(1-\rho^2)	\right]}{2(1+P_2T(1-\rho^2))} \nonumber \\
&\qquad+\frac{\left[-P_2T(1-\rho)^2\sum_{k=1}^n Z_k^2 + 2\sum_{k=1}^n Z_k\left(X_{2k}-\rho_k X_{1k}\sqrt{\frac{P_{2k}}{P_{1k}}}\right)+nP_2T(1-\rho^2)\right]}{2(1+P_2T(1-\rho^2))},
\end{align}
and
\begin{align}
&\sum_{k=1}^n \ln \frac{W(Y_k|X_{1k} X_{2k})}{Q_{Y_k}(Y_k)}\nn\\*
&=\frac{1}{2}\ln(1+P_1T+P_2T+2\rho\sqrt{P_1TP_2T})+\frac{\sum_{k=1}^nY_k^2-(1+P_1T+P_2T+2\rho T\sqrt{P_1P_2})\sum_{k=1}^n Z_k^2}{2(1+P_1T+P_2T+2\rho T\sqrt{P_1P_2})}\nonumber\\
&=\frac{1}{2}\ln(1+P_1T+P_2T+2\rho T\sqrt{P_1P_2})+\frac{[\sum_{k=1}^n(X_{1k}+X_{2k})^2-n(P_1T+P_2T+2\rho T\sqrt{P_1P_2})]}{2(1+P_1 T+P_2T+2\rho T \sqrt{P_1P_2})}\nonumber\\
&\qquad +\frac{[2\sum_{k=1}^n (X_{1k}+X_{2k}) Z_k -(P_1 T+P_2 T+2\rho T\sqrt{P_1P_2})\sum_{k=1}^n Z_k^2+n(P_1 T+P_2 T+2\rho T\sqrt{P_1P_2})]}{2(1+P_1 T+P_2 T+2\rho T\sqrt{P_1P_2})}.
\end{align}
Hence, we have
\begin{align}
&\Pr\left[\sum_{k=1}^n \ln \frac{W(Y_k|X_{1k} X_{2k})}{Q_{Y_k|X_{2k}}(Y_k|X_{2k})} \geq \ln\gamma_{1n}\right]\nn\\*
&=\Pr\left[\sum_{k=1}^n \ln \frac{W(Y_k|X_{1k} X_{2k})}{Q_{Y_k|X_{2k}}(Y_k|X_{2k})} \geq \ln\gamma_{1n}\Big| \mathcal{U}_{1n}^c\right] \Pr(\mathcal{U}_{1n}^c) +\Pr\left[\sum_{k=1}^n \ln \frac{W(Y_k|X_{1k} X_{2k})}{Q_{Y_k|X_{2k}}(Y_k|X_{2k})} \geq \ln\gamma_{1n}\Big| \mathcal{U}_{1n}\right] \Pr(\mathcal{U}_{1n})\\
&\leq \Pr(\mathcal{U}_{1n}^c) + \Pr\left[\sum_{k=1}^n \ln \frac{W(Y_k|X_{1k} X_{2k})}{Q_{Y_k|X_{2k}}(Y_k|X_{2k})} \geq \ln\gamma_{1n}\Big| \mathcal{U}_{1n}\right] \Pr(\mathcal{U}_{1n})\\
&=\Pr(\mathcal{U}_{1n}^c) + \Pr\left[\frac{V_{1n}}{2(1+P_1T(1-\rho^2))}\geq \ln\gamma_{1n}-\frac{n}{2} \ln\left[1+P_1T(1-\rho^2)\right] \right]\\
&=\Pr(\mathcal{U}_{1n}^c) + \Pr\left[\frac{V_{1n}}{2(1+P_1T(1-\rho^2))}\geq n^{2/3}\right] \\
&\stackrel{(a)}{\leq}\Pr(\mathcal{U}_{1n}^c)+\frac{\bbE[V_{1n}^2]}{[2(1+P_1T(1-\rho^2))]^2 n^{4/3}}\\
&\stackrel{(b)}{\leq} \frac{1}{T}+ \frac{2nP_1^2 T^2(1-\rho^2)^2 + 4 \sum_{k=1}^n P_{1k}(1-\rho_k^2)}{[2(1+P_1T(1-\rho^2))]^2 n^{4/3}} \\
&\stackrel{(c)}{\leq} \frac{1}{T}+ \frac{2nP_1^2 T^2(1-\rho^2)^2 + 4 n P_1(1-\rho^2)}{[2(1+P_1T(1-\rho^2))]^2 n^{4/3}}\\
&=\frac{1}{T}+O(n^{-1/3}).\label{eqn:finalO}
\end{align}
Here, (a) follows from Chebyshev's inequality \cite[Prop.~1.2]{TanBook}, (b) follows from~\eqref{mac-key-point2} and~\eqref{eq:eq163}, and (c) follows from~\eqref{eq:eq124}. Note that the implied constant in the $O(\cdot)$ notation in \eqref{eqn:finalO} depends only on $(P_1,P_2,\eps)$ since $1/(1-\eps)\le T \le {2}/{(1-\eps)}$ and $|\rho |\le 1$.

Using the same arguments, we can show that
\begin{align}
\Pr\left[\sum_{k=1}^n \ln \frac{W(Y_k|X_{1k} X_{2k})}{Q_{Y_k|X_{1k}}(Y_k|X_{1k})}\geq \ln \gamma_{2n}\right]\leq \frac{1}{T}+O(n^{-1/3}).
\end{align}

Finally, observe that
\begin{align}
&\Pr\left[\sum_{k=1}^n \ln \frac{W(Y_k|X_{1k} X_{2k})}{Q_{Y_k}(Y_k)}\geq \ln \gamma_{3n} \right]\nn\\*
&=\Pr\left[\sum_{k=1}^n \ln \frac{W(Y_k|X_{1k} X_{2k})}{Q_{Y_k}(Y_k)}\geq \ln \gamma_{3n}\Big|\mathcal{U}_{3n}^c\right] \Pr(\mathcal{U}_{3n}^c)+\Pr\left[\sum_{k=1}^n \ln \frac{W(Y_k|X_{1k} X_{2k})}{Q_{Y_k}(Y_k)}\geq \ln\gamma_{3n}\Big|\mathcal{U}_{3n}\right] \Pr(\mathcal{U}_{3n})\\
&\leq \Pr(\mathcal{U}_{3n}^c)+\Pr\left[\frac{V_{3n}}{1+P_1 T+P_2 T+2\rho\sqrt{P_1P_2}}\geq \ln \gamma_{3n}-\frac{n}{2}\ln(1+P_1+P_2+2\rho\sqrt{P_1P_2})\right]\\
&=\Pr(\mathcal{U}_{3n}^c)+\Pr\left[\frac{V_{3n}}{1+P_1 T+P_2 T+2\rho\sqrt{P_1P_2}}\geq  n^{2/3}\right]\\
&\stackrel{(a)}{\leq}\frac{1}{T} +\frac{\bbE[V_{3n}^2]}{\left(1+P_1 T+P_2 T+2\rho\sqrt{P_1P_2}\right)^2n^{4/3}}\\
&\stackrel{(b)}{=}\frac{1}{T}+\frac{4\sum_{k=1}^n (P_{1k}+P_{2k}+2\rho_k\sqrt{P_{1k}P_{2k}})+2n(P_1T+P_2T+2\rho T\sqrt{P_1P_2})^2}{\left(1+P_1 T+P_2 T+2\rho\sqrt{P_1P_2}\right)^2n^{4/3}}\\
&\stackrel{(c)}{\leq}\frac{1}{T}+\frac{4n (P_1+P_2+2\rho \sqrt{P_1P_2})+2n(P_1T+P_2T+2\rho T\sqrt{P_1P_2})^2}{\left(1+P_1 T+P_2 T+2\rho\sqrt{P_1P_2}\right)^2n^{4/3}}\\
&=\frac{1}{T}+O(n^{-1/3}).
\end{align}
Here, (a) follows from   Chebyshev's inequality \cite[Prop.~1.2]{TanBook} and~\eqref{mac-key-point2}, (b) follows from~\eqref{eq:eq165}, and (c) follows from~\eqref{eq:eq126}.
\end{IEEEproof}
\begin{proposition}
\label{mac-lem11}
For every sequence of $(n,M_{1n},M_{2n},\eps)$-codes for the AWGN-MAC with feedback under the expected power constraint, the constraints in \eqref{eqn:m_con1}--\eqref{eqn:m_con3} hold
for some $\rho \in [0,1]$. As a result, the information $\eps$-capacity region for the AWGN-MAC with feedback under an expected power constraint contains the  $\eps$-capacity region, i.e.,
\begin{align}
\mathcal{C}_{\mathrm{fb}}^*(P_1,P_2,\eps)  \subset  \overline{\mathcal{C}}_{\mathrm{fb}}^*(P_1,P_2,\eps).
\end{align}
\end{proposition}
\begin{IEEEproof}
Let the implied constants in the $O(\cdot)$ notation in \eqref{eqn:O1}--\eqref{eqn:O3} be $c_1=c_1(P_1,P_2,\eps), c_2=c_2(P_1,P_2,\eps)$, and  $c_{12}=c_{12}(P_1,P_2,\eps)$ respectively. From Lemmas~\ref{mac-lem9} and~\ref{mac-lem10}, we have the following inequalities for any $(n,M_{1n},M_{2n},\eps)$-code  for the AWGN-MAC with feedback:
\begin{align}
\label{eq:eq234}
\ln M_{1n} &\leq \frac{n}{2} \ln\left[1+P_1T(1-\rho^2)\right]+n^{2/3}-\ln^{+}\left(1-\eps-\frac{1}{T}-c_1n^{-1/3} \right),\\
\label{eq:eq235}
\ln M_{2n} &\leq \frac{n}{2} \ln\left[1+P_2T(1-\rho^2)\right]+n^{2/3}-\ln^{+}\left(1-\eps-\frac{1}{T}-c_2n^{-1/3} \right),\\
\label{eq:eq236}
\ln (M_{1n}M_{2n}) &\leq \frac{n}{2}\ln\left[1+P_1T+P_2T+2\rho T\sqrt{P_1P_2}\right]+ n^{2/3}-\ln^{+}\left(1-\eps-\frac{1}{T}-c_{12} n^{-1/3} \right).
\end{align}  for any positive real number $1/(1-\eps) \le T\le  2/(1-\eps)$.

Let $c_{\max}:=\max\{c_1,c_2,c_{12}\}$. Now we set
\begin{align}
\label{eq:eq237}
T:=\frac{1}{1-\eps-(c_{\max}+1) n^{-1/3}}.
\end{align}  Note that  the value of $T$ is in $[ 1/{( 1-\eps) },2/{( 1-\eps) }]$ for $n$ sufficiently large (depending only on $P_1,P_2$ and $\eps$) so Lemma~\ref{mac-lem10} readily applies. With this choice of $T$, all the $\ln^+(\cdot )$ terms in \eqref{eq:eq234}--\eqref{eq:eq236} are $O(\ln n)$.

Thus from~\eqref{eq:eq234}--\eqref{eq:eq236}, we have
\begin{align}
\ln M_{1n} &\leq \frac{n}{2}\ln\left[1+\frac{P_1(1-\rho^2)}{1-\eps-(c_{\max}+1) n^{-1/3}}\right] +O(n^{2/3}),\\
&\stackrel{(a)}{=}\frac{n}{2}\ln\left[1+\frac{P_1(1-\rho^2)}{1-\eps}\right] +O(n^{2/3}).
\end{align} Here, (a) follows from   Taylor's expansion.

Similarly, we can show that
\begin{align}
\ln M_{2n} \leq \ln\left[1+\frac{P_2(1-\rho^2)}{1-\eps}\right] +O(n^{2/3}),
\end{align}
and
\begin{align}
\ln(M_{1n}M_{2n})&\leq \frac{n}{2}\ln\left[1+\frac{P_1+P_2+2\rho\sqrt{P_1P_2}}{1-\eps}\right] + O(n^{2/3})
\end{align}
Since $|\rho|\le 1$ (by Lemma~\ref{mac-lem6}), it follows that the  $\eps$-capacity region  satisfies
\begin{align}
\mathcal{C}_{\mathrm{fb}}^*(P_1,P_2,\eps)\subset \bigcup_{-1\leq \rho\leq 1}\calC(\rho; P_1, P_2,\eps)=\bigcup_{0\leq \rho\leq 1}\calC(\rho; P_1, P_2,\eps)=\overline{\mathcal{C}}_{\mathrm{fb}}^*(P_1,P_2,\eps)
\end{align}
completing the proof of Proposition \ref{mac-lem11}, and hence the converse proof of Theorem \ref{mac-thm4}.
\end{IEEEproof}
\section{Conclusion} \label{sec:concl}
In this paper, we     have made some   progress in bounding the maximum rate of transmission over an AWGN channel with  feedback and an expected power constraint and with a non-vanishing error probability. We    have also found the $\eps$-capacity region for the AWGN-MAC with feedback under the same settings (constraints) as the AWGN channel. For both channel models, we have established bounds on the second-order asymptotics.

It would be fruitful, though challenging, to derive the exact second-order coding region for both problems. A less challenging endeavor is to tighten the order of the second-order remainder terms for the direct and converse parts. Another interesting direction is to  establish the $\eps$-capacity regions for other multi-terminal channel models with feedback such as the  Gaussian broadcast channel~\cite[Example 17.3]{elgamal}, the relay channel~\cite[Section 17.4]{elgamal}, or the  two-way channel~\cite[Section 17.5]{elgamal}.

\appendices

\section{Proof of Lemma \ref{mac-lem20}}\label{app:prf_continuity}
By \eqref{eqn:diff_coeffs}, each of  the coefficients $b_j$ differs from $a_j$ by no more than $d/n$, where $0<d<\infty$ is a constant. The solutions to any quartic equation are known in closed-form  \cite[Sec.~3.8.3]{Abramowitz} and are continuously differentiable functions of the coefficients (containing surds, polynomials, etc.). Thus, the solutions to  \eqref{eqn:a_quartic} and \eqref{eqn:b_quartic} are continuously differentiable functions of the coefficients. Let us call the function that maps the coefficients $\ba=(a_0,a_1,\ldots, a_4)$ as to the root $z^*$ as $f(\ba)$. Now we employ a Taylor expansion which asserts that for any continuously differentiable function $f(\ba)$, we have
\begin{equation}
 f\bigg(\ba + O\Big(\frac{1}{n}\Big) \bone\bigg)  =f(\ba)+O\bigg( \frac{1}{n} \bigg), \label{eqn:tayl}
\end{equation}
where $\bone$ is the all-ones vector.
  Now we note that the left-hand-side $f(\ba+O(1/n)\bone)$ produces the solution to the quartic with perturbed coefficients in~\eqref{eqn:b_quartic}. From \eqref{eqn:tayl}, we see that
there exists a  solution to \eqref{eqn:b_quartic}, namely $z_n^*$,  that is  of the order $1/n$ away from the solution to \eqref{eqn:a_quartic}, namely $z^*$.
\section{Proof of Lemma \ref{mac-lem8} }\label{app:prf_moments}

Observe that
\begin{align}
\bbE[V_{1n}]=-P_1T(1-\rho)^2\sum_{k=1}^n \bbE[Z_k^2] + 2\sum_{k=1}^n \bbE[Z_k]\bbE\left[\left(X_{1k}-\rho_k X_{2k}\sqrt{\frac{P_{1k}}{P_{2k}}}\right)\right]+nP_1T(1-\rho^2)=0.
\end{align}
Moreover, we also have
\begin{align}
\label{eq:eq167}
\bbE[V_{1n}^2]&=[P_1T(1-\rho^2)]^2 \bbE\left[\left(\sum_{k=1}^n Z_k^2\right)^2\right] + 4\bbE\left[\sum_{k=1}^n Z_k \left(X_{1k}-\rho_k X_{2k}\sqrt{\frac{P_{1k}}{P_{2k}}}\right)\right]^2+[nP_1T(1-\rho^2)]^2 \nonumber \\
&\qquad -4P_1T(1-\rho^2)\bbE\left[\left(\sum_{k=1}^n Z_k^2\right)\left(\sum_{k=1}^n Z_k(X_{1k}-\rho_k X_{2k}\sqrt{\frac{P_{1k}}{P_{2k}}}\right)\right]-2nP_1^2T^2(1-\rho^2)^2\bbE\left[\sum_{k=1}^n Z_k^2\right]\nonumber\\
&\qquad +2nP_1T(1-\rho^2)\sum_{k=1}^n \bbE\left[Z_k\left(X_{1k}-\rho_k X_{2k}\sqrt{\frac{P_{1k}}{P_{2k}}}\right)\right].
\end{align}
Now, observe that
\begin{align}
\bbE\left[\left(\sum_{k=1}^n Z_k^2\right)^2\right]&=\bbE\left[\sum_{k=1}^n Z_k^4 + 2\sum_{1\leq i<j\leq n} Z_i^2 Z_j^2\right]\\
&=\sum_{k=1}^n \bbE[Z_k^4] +2 \sum_{1\leq i<j\leq n} \bbE[Z_i^2] \bbE[Z_j^2]\\
&=3n+n(n-1)=n^2 + 2n. \label{eq:eq170}
\end{align}
Furthermore,
\begin{align}
&\bbE\left[\sum_{k=1}^n Z_k \left(X_{1k}-\rho_k X_{2k}\sqrt{\frac{P_{1k}}{P_{2k}}}\right)\right]^2\nn\\*
&=\bbE\left[\sum_{k=1}^n Z_k^2 \left(X_{1k}-\rho_k X_{2k}\sqrt{\frac{P_{1k}}{P_{2k}}}\right)^2\right] +2\bbE\left[\sum_{1\leq i<j\leq n}Z_i\left(X_{1i}-\rho_i X_{2i}\sqrt{\frac{P_{1i}}{P_{2i}}}\right) Z_j\left(X_{1j}-\rho_j X_{2j}\sqrt{\frac{P_{1j}}{P_{2j}}}\right)\right] \\
&\stackrel{(a)}{=}\sum_{k=1}^n \bbE[Z_k^2] \bbE\left[\left(X_{1k}-\rho_k X_{2k}\sqrt{\frac{P_{1k}}{P_{2k}}}\right)^2\right] \nn\\*
&\qquad  + 2\sum_{1\leq i<j\leq n}\bbE\left[Z_i\left(X_{1i}-\rho_i X_{2i}\sqrt{\frac{P_{1i}}{P_{2i}}}\right) \left(X_{1j}-\rho_j X_{2j}\sqrt{\frac{P_{1j}}{P_{2j}}}\right)\right]\bbE[Z_j] \\
&=\sum_{k=1}^n \bbE\left[\left(X_{1k}-\rho_k X_{2k}\sqrt{\frac{P_{1k}}{P_{2k}}}\right)^2\right] \\
&=\sum_{k=1}^n P_{1k}(1-\rho_k^2),\label{eq:eq174}
\end{align}
where (a) follows from the fact that $Z_j$ is independent of $(X_{1i}, X_{2i}, Z_i)$ for $j>i$.

Next, we have
\begin{align}
&\bbE\left[\left(\sum_{k=1}^n Z_k^2\right)\left(\sum_{k=1}^n Z_k(X_{1k}-\rho_k X_{2k}\sqrt{\frac{P_{1k}}{P_{2k}}}\right)\right]\nn\\*
&=\bbE\left[\sum_{k=1}^n Z_k^3 \left(X_{1k}-\rho_k X_{2k}\sqrt{\frac{P_{1k}}{P_{2k}}}\right)\right]\nonumber\\*
&\qquad +\sum_{1\leq i<j \leq n}\bbE\left[Z_i^2 Z_j\left(X_{1j}-\rho_j X_{2j}\sqrt{\frac{P_{1j}}{P_{2j}}}\right)\right]\nonumber\\*
&\qquad +\sum_{1\leq j<i \leq n}\bbE\left[Z_i^2 Z_j\left(X_{1j}-\rho_j X_{2j}\sqrt{\frac{P_{1j}}{P_{2j}}}\right)\right] \\
&\stackrel{(a)}{=}  \sum_{k=1}^n \bbE[Z_k^3] \bbE\left(X_{1k}-\rho_k X_{2k}\sqrt{\frac{P_{1k}}{P_{2k}}}\right)\nonumber\\*
&\qquad +\sum_{1\leq i<j \leq n}\bbE\left[Z_i^2 \left(X_{1j}-\rho_j X_{2j}\sqrt{\frac{P_{1j}}{P_{2j}}}\right)\right]\bbE[Z_j]\nonumber\\*
&\qquad +\sum_{1\leq j<i \leq n}\bbE\left[ Z_j\left(X_{1j}-\rho_j X_{2j}\sqrt{\frac{P_{1j}}{P_{2j}}}\right)\right]\bbE[Z_i^2] \\
&\stackrel{(b)}{=} \sum_{k=1}^n \bbE[Z_k^3] \bbE\left(X_{1k}-\rho_k X_{2k}\sqrt{\frac{P_{1k}}{P_{2k}}}\right)\nonumber\\*
&\qquad +\sum_{1\leq i<j \leq n}\bbE\left[Z_i^2 \left(X_{1j}-\rho_j X_{2j}\sqrt{\frac{P_{1j}}{P_{2j}}}\right)\right]\bbE[Z_j]\nonumber\\*
&\qquad +\sum_{1\leq j<i \leq n}\bbE[Z_j] \bbE\left[ \left(X_{1j}-\rho_j X_{2j}\sqrt{\frac{P_{1j}}{P_{2j}}}\right)\right]\bbE[Z_i^2]\\
\label{eq:eq178}
&=0+0+0=0.
\end{align}
Here, (a) follows from the fact that $Z_j$ is independent of $(X_{1i}, X_{2i}, Z_i)$ for $j>i$, and (b) follows from the fact that $Z_j$ is independent of $\left(X_{1j}-\rho_j X_{2j}\sqrt{\frac{P_{1j}}{P_{2j}}}\right)$.

Now consider
\begin{align}
\label{eq:eq179}
\bbE\left[\sum_{k=1}^n Z_k^2\right]&=\sum_{k=1}^n \bbE[Z_k^2]=n,\\
\label{eq:eq180}
\bbE\left[Z_k\left(X_{1k}-\rho_k X_{2k}\sqrt{\frac{P_{1k}}{P_{2k}}}\right)\right]&=\bbE[Z_k]\bbE\left[\left(X_{1k}-\rho_k X_{2k}\sqrt{\frac{P_{1k}}{P_{2k}}}\right)\right]=0.
\end{align}
Substituting~\eqref{eq:eq170},~\eqref{eq:eq174},~\eqref{eq:eq178},~\eqref{eq:eq179},~\eqref{eq:eq180} into~\eqref{eq:eq167} we obtain
\begin{align}
\label{eq:eq181}
\bbE[V_{1n}^2]&=[P_1T(1-\rho^2)]^2 (n^2 +2n)+4\sum_{k=1}^n P_{1k}(1-\rho_k^2)+[nP_1T(1-\rho^2)]^2\nonumber\\
&\qquad -4P_1 T (1-\rho^2) \times 0 -2n^2P_1^2 T^2 (1-\rho^2)^2+2nP_1 T(1-\rho^2) \times 0  \nonumber \\
&=2nP_1^2 T^2(1-\rho^2)^2 + 4 \sum_{k=1}^n P_{1k}(1-\rho_k^2).
\end{align}
Similarly, we have
\begin{align}
\bbE[V_{2n}]&=0,\\
\bbE[V_{2n}^2]&=2nP_2^2 T^2(1-\rho^2)^2 + 4 \sum_{k=1}^n P_{2k}(1-\rho_k^2).
\end{align}
Finally, note that
\begin{align}
\bbE[V_{3n}]&=2\sum_{k=1}^n \bbE[X_{1k}+X_{2k}] \bbE[Z_k ]-(P_1 T+P_2 T+2\rho T\sqrt{P_1P_2})\sum_{k=1}^n \bbE[Z_k^2]+n(P_1 T+P_2 T+2\rho T\sqrt{P_1P_2})=0,\\
\bbE[V_{3n}^2]&=4\bbE\left[\sum_{k=1}^n (X_{1k}+X_{2k})Z_k \right]^2+(P_1 T+P_2T +2\rho T \sqrt{P_1 P_2})^2\bbE\left[\sum_{k=1}^n Z_k^2\right]^2\nonumber\\
&\qquad + n^2 (P_1 T+P_2 T+2\rho T\sqrt{P_1P_2})^2 - 4(P_1T+P_2T+2\rho\sqrt{P_1P_2})\bbE\left[\left(\sum_{k=1}^n(X_{1k}+X_{2k})Z_k\right)\left(\sum_{k=1}^n Z_k^2\right)\right]\nonumber\\
&\qquad +4n(P_1T +P_2 T+2\rho T\sqrt{P_1P_2})\bbE\left[\sum_{k=1}^n(X_{1k}+X_{2k})Z_k\right]
\label{eq:eq185} \nn\\*
&\qquad -2n(P_1T+P_2T+2\rho T\sqrt{P_1P_2})^2\bbE\left[\sum_{k=1}^n Z_k^2\right].
\end{align}
Now, we see that
\begin{align}
&\bbE\left[\sum_{k=1}^n (X_{1k}+X_{2k})Z_k \right]^2\nn\\*
&=\sum_{k=1}^n \bbE[(X_{1k}+X_{2k})^2]\bbE[Z_k^2]+ 2\sum_{1\leq i<j\leq n} \bbE[(X_{1i}+X_{2i})Z_i (X_{1j}+X_{2j})Z_j]\\
\label{eq:eq187}
&=\sum_{k=1}^n \bbE[(X_{1k}+X_{2k})^2]+2\sum_{1\leq i<j\leq n} \bbE[(X_{1i}+X_{2i})Z_i (X_{1j}+X_{2j})]\bbE[Z_j]\\
\label{eq:eq188}
&=\sum_{k=1}^n P_{1k}+P_{2k}+2\rho_k \sqrt{P_{1k}P_{2k}},
\end{align}
and
\begin{align}
\bbE\left[\left(\sum_{k=1}^n Z_k^2\right)^2\right] &= n^2+2n.\label{eq:eq189}
\end{align}In addition,
\begin{align}
&\bbE\left[\left(\sum_{k=1}^n(X_{1k}+X_{2k})Z_k\right)\left(\sum_{k=1}^n Z_k^2\right)\right]\nn\\*
&=\sum_{k=1}^n \bbE\left[(X_{1k}+X_{2k})\right]\bbE[Z_k^3]\nonumber \\
&\qquad +\sum_{1\leq i<j \leq n} \bbE[(X_{1i}+X_{2i})Z_i Z_j^2] +\sum_{1\leq i<j \leq n} \bbE[(X_{1j}+X_{2j})Z_j Z_i^2]\\
&=\sum_{k=1}^n \bbE\left[(X_{1k}+X_{2k})\right]\bbE[Z_k^3]\nonumber \\
&\qquad +\sum_{1\leq i<j \leq n} \bbE[(X_{1i}+X_{2i})]\bbE[Z_i]\bbE[Z_j^2] +\sum_{1\leq i<j \leq n} \bbE[(X_{1j}+X_{2j}) Z_i^2]\bbE[Z_j]\\
\label{eq:eq192}
&=0+0+0=0.
\end{align}
Substituting~\eqref{eq:eq188},~\eqref{eq:eq189},~\eqref{eq:eq192} into \eqref{eq:eq185} we obtain
\begin{align}
\label{eq:eq193}
\bbE[V_{3n}^2]&=4\sum_{k=1}^n (P_{1k}+P_{2k}+2\rho_k\sqrt{P_{1k}P_{2k}})+(P_1T+P_2T+2\rho T\sqrt{P_1P_2})^2 (n^2+2n)\nonumber\\
&\qquad +n^2(P_1T+P_2T+2\rho T\sqrt{P_1P_2})^2-2n^2(P_1T+P_2T+2\rho T\sqrt{P_1P_2})^2 \nonumber \\
&=4\sum_{k=1}^n (P_{1k}+P_{2k}+2\rho_k\sqrt{P_{1k}P_{2k}})+2n(P_1T+P_2T+2\rho T\sqrt{P_1P_2})^2
\end{align}
as desired.
\end{IEEEproof}

\section{Proof of Lemma \ref{mac-lem9}} \label{app:info_spec}
In the proof of Lemma \ref{mac-lem9}, we use the following result concerning the non-asymptotic fundamental limits of binary hypothesis testing.
\begin{lemma}
Consider a set $\mathcal{X}$ and two distributions $P$ and $Q$ on $\calX$. Let $\beta_\alpha(P,Q)$ be the smallest type-II error probability of a (randomized) binary hypothesis test $H\in\{0,1\}$ between $P$ and $Q$ with the type-I error probability being no larger than $1-\alpha$, i.e.,
\begin{equation}
\beta_\alpha(P,Q):=\min_{ \substack{P_{H|X}:\calX\to\{0,1\} :\\ \sum_{x\in\calX} P_{H|X}(1 | x) P(x) \ge\alpha}} \sum_{x\in\calX}P_{H|X}(1|x)Q(x).
\end{equation}
 Then, the following two statements hold:
\begin{itemize}
\item Data Processing Inequality (DPI)
\begin{align}
\label{eq:eq245}
\beta_{\alpha}(P,Q) \leq \beta_{\alpha}(PV,QV)
\end{align} for any channel $V$ from $\calX$ to another set $\calY$ and $PV(y)=\sum_x V(y|x) P(x)$ is the output distribution induced by $P$ and $V$.
\item  For any $\eta>0$
\begin{align}
\label{eq:eq246}
\beta_{\alpha}(P,Q)\geq \frac{1}{\eta}\left(\alpha-P\left[\ln \frac{P}{Q}\geq \ln \eta\right]\right),
\end{align}
\end{itemize}
\end{lemma}
The proofs of these results can be found in \cite[Sec.~2.3]{Pol10} and \cite[Sec.~2.1]{TanBook}.

Now, we use similar arguments as in~\cite{FT15},  \cite{Pol10},    and \cite{FT14} to prove Lemma~\ref{mac-lem6}. Encoder $1$ defines a transition probability kernel $P_{Y^n|W_1}(y^n|w_1)$ from the input space $\mathcal{D}_{1}:=\{1,2,\ldots,M_{1n} \}$ to the output  $Y^n$. Hence, we can view the triplet $(\calD_{1},P_{Y^n|W_1} ,\bbR^n)$ as a random transformation for which we can use the meta-converse theorem~\cite[Section III.E]{PPV10}. Note the error probability here is bounded as $\Pr(\hat{W}_1\ne W_1)\le\Pr((\hat{W}_1,\hat{W}_2)\neq (W_1,W_2)) \leq \eps$. Therefore, we have
\begin{align}
\label{eq:eq250}
M_{1n}\leq& \frac{1}{\beta_{1-\eps}\left(P_{W_1Y^n},P_{W_1} Q_{Y^n}\right)},
\end{align}
where $P_{W_1}$ is equiprobable on $\calD_1$ and the inequality holds
for any auxiliary output distribution $Q_{Y^n}$. Define
\begin{align}
\label{eq:eq243a}
Q_{X_2^n Y^n|W_2}&:=\prod_{k=1}^n P_{X_{2k}|W_2,Y^{k-1}}Q_{Y_k|X_{2k}},\\
\label{eq:eq243b}
Q_{W_2 X_2^n Y^n}&:=P_{W_2}Q_{X_2^nY^n|W_2}.
\end{align} In the rest of the proof, we use $W$ or $W_{Y_k | X_{1k} X_{2k}}$ to denote the $k$-th channel. Note that by stationarity, all these channels are the same but we sometimes retain the time index $k$ for the sake of clarity.

Applying the DPI in~\eqref{eq:eq245}, we see that
\begin{align}
&\beta_{1-\eps}(P_{W_1Y^n},P_{W_1}Q_{Y^n}) \nn\\*
&\geq \beta_{1-\eps}\left(P_{W_1W_2Y^n},P_{W_1} Q_{W_2Y^n}\right)\\*
&\stackrel{(a)}{=}\beta_{1-\eps}\left(P_{W_1W_2} P_{Y^n|W_1W_2},P_{W_1}P_{W_2} Q_{Y^n|W_2}\right)\\
&=\beta_{1-\eps}\left(P_{W_1W_2} P_{Y^n|W_1W_2},P_{W_1W_2} Q_{Y^n|W_2}\right)\\
&\geq \beta_{1-\eps}\left(P_{W_1W_2} P_{X_1^n X_2^n Y^n|W_1W_2},P_{W_1W_2} Q_{X_2^n Y^n|W_2} \prod_{k=1}^n P_{X_{1k}|W_1W_2X_1^{k-1}X_2^{k-1}Y^{k-1}}\right)\\
&= \beta_{1-\eps}\left(P_{W_1W_2} \prod_{k=1}^n P_{X_{1k} X_{2k}|W_1W_2X_1^{k-1}X_2^{k-1}Y^{k-1}}\prod_{k=1}^n W_{Y_k|X_{1k}X_{2k}},\right.\nonumber\\
&\qquad\qquad\qquad \left. P_{W_1W_2} Q_{X_2^n Y^n|W_2}\prod_{k=1}^n P_{X_{1k}|W_1W_2X_1^{k-1}X_2^{k-1}Y^{k-1}}\right)  \\
&\stackrel{(b)}{=} \beta_{1-\eps}\left(P_{W_1W_2} \prod_{k=1}^n P_{X_{1k} X_{2k}|W_1W_2X_1^{k-1}X_2^{k-1}Y^{k-1}}\prod_{k=1}^n W_{Y_k|X_{1k}X_{2k}},\right.\nonumber\\
&\qquad\qquad\qquad\left. P_{W_1W_2} \prod_{k=1}^n P_{X_{2k}|W_2,Y^{k-1}}\prod_{k=1}^n Q_{Y_k|X_{2k}}\prod_{k=1}^n P_{X_{1k}|W_1W_2X_1^{k-1}X_2^{k-1}Y^{k-1}}\right)  \\
&\stackrel{(c)}{=} \beta_{1-\eps}\left(P_{W_1W_2} \prod_{k=1}^n P_{X_{1k}|W_1W_2X_1^{k-1}X_2^{k-1}Y^{k-1}}\prod_{k=1}^n P_{X_{2k}|W_2Y^{k-1}}\prod_{k=1}^n W_{Y_k|X_{1k}X_{2k}}\right., \nonumber \\*
\label{eq:eq256}
&\qquad\qquad\qquad\left. P_{W_1W_2} \prod_{k=1}^n P_{X_{2k}|W_2,Y^{k-1}}\prod_{k=1}^n Q_{Y_k|X_{2k}}\prod_{k=1}^n P_{X_{1k}|W_1W_2X_1^{k-1}X_2^{k-1}Y^{k-1}}\right).
\end{align}
Here, (a), (b) follow from~\eqref{eq:eq243a} and~\eqref{eq:eq243b} and (c) follows from the Markov chain
\begin{equation}
(W_1,X_1^k, X_2^{k-1})-(W_2 ,Y^{k-1})-X_{2k}.
\end{equation}

By using \eqref{eq:eq246}, from \eqref{eq:eq250} and \eqref{eq:eq256} we have for any $\gamma_{1n}>0$ and any   sequence of auxiliary  distributions  $Q_{Y_k|X_{2k}},k=1,2,\ldots,n$ that
\begin{align}
\frac{1}{M_{1n}} \geq \frac{1}{\gamma_{1n}}\left(1-\eps-\Pr\left[\sum_{k=1}^n \ln \frac{W(Y_k|X_{1k} X_{2k})}{Q_{Y_k|X_{2k}}(Y_k|X_{2k})}\geq \ln \gamma_{1n}\right]\right).
\end{align}
Note here that we exploited the fact that the MAC is stationary and memoryless so $W^n(y^n|x_1^n,x_2^n)=\prod_{k=1}^nW(y_k|x_{1k},x_{2k})$.
It follows that
\begin{align}
\ln M_{1n}\leq \ln \gamma_{1n} -\ln^{+}\left(1-\eps-\Pr\left[\sum_{k=1}^n \ln \frac{W(Y_k|X_{1k} X_{2k})}{Q_{Y_k|X_{2k}}(Y_k|X_{2k})}\geq \ln \gamma_{1n}\right]\right).
\end{align}
Similarly, we can show that for all $\gamma_{2n}>0$ and any   sequence of auxiliary  distributions  $Q_{Y_k|X_{1k}},k=1,2,\ldots,n$ that
\begin{align}
\ln M_{2n}\leq \ln \gamma_{2n} -\ln^{+}\left(1-\eps-\Pr\left[\sum_{k=1}^n \ln\frac{W(Y_k|X_{1k} X_{2k})}{Q_{Y_k|X_{1k}}(Y_k|X_{1k})}\geq \ln\gamma_{2n}\right]\right).
\end{align}
Now, the combination of the two encoders defines a transition probability kernel $P_{Y^n|W_1W_2}$ from an input space $\mathcal{D}_{1,2}:=\{1,2,\ldots ,M_{1n}\}\times \{1,2\ldots ,M_{2n}\}$ to the output $Y^n$. We can view then the triplet $(\mathcal{D}_{1,2}, P_{Y^n|W_1W_2},\mathbb{R}^n)$ as a random transformation for which we have a usual $(M_{1n}M_{2n},\eps)$-code in the sense of  \cite[Definition 2]{Pol10}  with error probability $\Pr((\hat{W}_1,\hat{W}_2)\neq (W_1,W_2)) \leq \eps$. For such a code, by the meta-converse theorem~\cite[Section III.E]{PPV10}
\begin{align}
\label{eq:eq260}
M_{1n} M_{2n} \leq \frac{1}{\beta_{1-\eps}(P_{W_1W_2}P_{Y^n|W_1W_2},P_{W_1W_2}Q_{Y^n})},
\end{align} where $P_{W_1W_2}$ is the equiprobable distribution on $\mathcal{D}_{1,2}$ and $Q_{Y^n}$ is arbitrary. Define
\begin{align}
Q_{Y^n}=\prod_{k=1}^n Q_{Y_k}
\end{align}
for some single-letter output distributions $Q_{Y_k}, k=1,2,\ldots, n$.
Using the DPI in~\eqref{eq:eq245}, we have
\begin{align}
&\beta_{1-\eps}\left(P_{W_1W_2}P_{Y^n|W_1W_2},P_{W_1W_2}Q_{Y^n}\right) \\
&\geq \beta_{1-\eps}\left(P_{W_1W_2}P_{X_1^nX_2^nY^n|W_1W_2},P_{W_1W_2} Q_{Y^n}\prod_{k=1}^n P_{X_{1k}X_{2k}|W_1W_2X_1^{k-1}X_2^{k-1} Y^{k-1}}\right) \\
&=\beta_{1-\eps}\left(P_{W_1W_2} \prod_{k=1}^n P_{X_{1k}X_{2k}Y_k|W_1W_2X_1^{k-1}X_2^{k-1}Y^{k-1}},P_{W_1W_2} Q_{Y^n}\prod_{k=1}^n P_{X_{1k}X_{2k}|W_1W_2X_1^{k-1}X_2^{k-1} Y^{k-1}}\right)\\
\label{eq:eq265}
&\stackrel{(a)}{=}\beta_{1-\eps}\Bigg(P_{W_1W_2} \prod_{k=1}^n P_{X_{1k}X_{2k}|W_1W_2X_1^{k-1}X_2^{k-1} Y^{k-1}} \prod_{k=1}^n W_{Y_k|X_{1k}X_{2k}},  \nn\\*
&\qquad \qquad\qquad P_{W_1W_2} Q_{Y^n}\prod_{k=1}^n P_{X_{1k}X_{2k}|W_1W_2X_1^{k-1}X_2^{k-1} Y^{k-1}}\Bigg),
\end{align}
where (a) follows from the Markov chain
\begin{equation}
(W_1,W_2, X_1^{k-1}, X_2^{k-1}, Y^{k-1})-(X_{1k},X_{2k})-Y_k.
\end{equation}

Now, applying~\eqref{eq:eq246}, from~\eqref{eq:eq260} and~\eqref{eq:eq265} we have for any $\gamma_{3n}>0$  and any sequence of auxiliary distributions $Q_{Y_k},k=1,2,\ldots,n$ that
\begin{align}
\frac{1}{M_{1n} M_{2n}} \geq \frac{1}{\gamma_{3n}}\left(1-\eps-\Pr\left[\sum_{k=1}^n \ln \frac{W(Y_k|X_{1k} X_{2k})}{Q_{Y_k}(Y_k)}\geq \ln \gamma_{3n}\right]\right).
\end{align}
This is equivalent to
\begin{align}
\ln \left(M_{1n} M_{2n}\right) \leq \ln \gamma_{3n} -\ln^{+}\left(1-\eps-\Pr\left[\sum_{k=1}^n \ln \frac{W(Y_k|X_{1k} X_{2k})}{Q_{Y_k}(Y_k)}\geq \ln \gamma_{3n}\right]\right),
\end{align}
completing the proof of Lemma \ref{mac-lem9}.


\subsection*{Acknowledgements}
Discussions with Dr.\ Wei Yang (Princeton) are gratefully acknowledged.

The authors would also like to thank the Associate Editor, Prof.\ Haim Permuter  and the reviewers for their careful reading and constructive suggestions. In particular, a reviewer suggested to adapt the coding schemes in \cite{GalN, Ihara12} to improve the achievable second-order term in \eqref{eqn:dire}  in Theorem \ref{thm:main}.

%
%
%
%

%
%
%
%
%




\end{document}

%% file: preamble.tex
\usepackage[mathscr]{eucal}
\usepackage[cmex10]{amsmath}
\usepackage{epsfig,epsf,psfrag}
\usepackage{amssymb,amsmath,amsthm,amsfonts,latexsym}
\usepackage{amsmath,graphicx,bm,xcolor,url}
\usepackage[caption=false]{subfig} 
\usepackage{fixltx2e}
\usepackage{array}
\usepackage{verbatim}
\usepackage{bm}
\usepackage{algorithmic, cite}
\usepackage{algorithm}
\usepackage{verbatim}
\usepackage{textcomp}
\usepackage{mathrsfs}
\usepackage{multirow}
\usepackage{epstopdf}

\catcode`~=11 \def\UrlSpecials{\do\~{\kern -.15em\lower .7ex\hbox{~}\kern .04em}} \catcode`~=13 

\allowdisplaybreaks[1]
 
\newcommand{\nn}{\nonumber}

\newcommand{\calA}{\mathcal{A}}

\newcommand{\calC}{\mathcal{C}}
\newcommand{\calD}{\mathcal{D}}
\newcommand{\calE}{\mathcal{E}}

\newcommand{\calN}{\mathcal{N}}

\newcommand{\calW}{\mathcal{W}}
\newcommand{\calX}{\mathcal{X}}
\newcommand{\calY}{\mathcal{Y}}

\newcommand{\ba}{\mathbf{a}}

\newcommand{\rmd}{\mathrm{d}}

\newcommand{\rme}{\mathrm{e}}

\newcommand{\rmQ}{\mathrm{Q}}

\newcommand{\bbE}{\mathbb{E}}

\newcommand{\bbN}{\mathbb{N}}

\newcommand{\bbR}{\mathbb{R}}

\DeclareMathAlphabet{\mathbsf}{OT1}{cmss}{bx}{n}
\DeclareMathAlphabet{\mathssf}{OT1}{cmss}{m}{sl}

\newcommand{\rvC}{\mathsf{C}}

\newcommand{\rvV}{\mathsf{V}}

\DeclareSymbolFont{bsfletters}{OT1}{cmss}{bx}{n}  
\DeclareSymbolFont{ssfletters}{OT1}{cmss}{m}{n}
\DeclareMathSymbol{\bsfGamma}{0}{bsfletters}{'000}
\DeclareMathSymbol{\ssfGamma}{0}{ssfletters}{'000}
\DeclareMathSymbol{\bsfDelta}{0}{bsfletters}{'001}
\DeclareMathSymbol{\ssfDelta}{0}{ssfletters}{'001}
\DeclareMathSymbol{\bsfTheta}{0}{bsfletters}{'002}
\DeclareMathSymbol{\ssfTheta}{0}{ssfletters}{'002}
\DeclareMathSymbol{\bsfLambda}{0}{bsfletters}{'003}
\DeclareMathSymbol{\ssfLambda}{0}{ssfletters}{'003}
\DeclareMathSymbol{\bsfXi}{0}{bsfletters}{'004}
\DeclareMathSymbol{\ssfXi}{0}{ssfletters}{'004}
\DeclareMathSymbol{\bsfPi}{0}{bsfletters}{'005}
\DeclareMathSymbol{\ssfPi}{0}{ssfletters}{'005}
\DeclareMathSymbol{\bsfSigma}{0}{bsfletters}{'006}
\DeclareMathSymbol{\ssfSigma}{0}{ssfletters}{'006}
\DeclareMathSymbol{\bsfUpsilon}{0}{bsfletters}{'007}
\DeclareMathSymbol{\ssfUpsilon}{0}{ssfletters}{'007}
\DeclareMathSymbol{\bsfPhi}{0}{bsfletters}{'010}
\DeclareMathSymbol{\ssfPhi}{0}{ssfletters}{'010}
\DeclareMathSymbol{\bsfPsi}{0}{bsfletters}{'011}
\DeclareMathSymbol{\ssfPsi}{0}{ssfletters}{'011}
\DeclareMathSymbol{\bsfOmega}{0}{bsfletters}{'012}
\DeclareMathSymbol{\ssfOmega}{0}{ssfletters}{'012}

\newcommand{\eps}{\varepsilon}

\DeclareMathOperator{\var}{\mathsf{Var}}

\newcommand{\bone}{\mathbf{1}}

\newtheorem{theorem}{Theorem} 
\newtheorem{lemma}[theorem]{Lemma}

\newtheorem{proposition}[theorem]{Proposition}

\newtheorem{definition}{Definition}

\newtheorem{remark}{Remark}

\newcommand{\qednew}{\nobreak \ifvmode \relax \else
      \ifdim\lastskip<1.5em \hskip-\lastskip
      \hskip1.5em plus0em minus0.5em \fi \nobreak
      \vrule height0.75em width0.5em depth0.25em\fi}